\documentclass{aa}  

\def\jobname{aa59935-26}

\usepackage[version=3]{mhchem}
\usepackage{color}
\definecolor{darkgreen}{rgb}{0,0.7,0}
\usepackage{graphicx}
\usepackage{gensymb}
\usepackage{hyperref}
\usepackage{float}
\usepackage{here}
\usepackage{subfig}
\usepackage{threeparttable}
\hypersetup{
    colorlinks=true,
    citecolor=blue,
    linkcolor=red,
    urlcolor=black
    }  

\usepackage{txfonts}

\begin{document}

   \title{Modeling the chemical impacts of luminosity outbursts in protostellar envelopes}

   \author{A. Espagnet\inst{1}
   \and A. Coutens\inst{1}
   \and P. Marchand\inst{1}
   \and J.-C. Loison\inst{2}
   \and F. Cruz-Sáenz de Miera\inst{1}
   \and V. Wakelam\inst{3}
   }
   
   \institute{Univ Toulouse, CNES, CNRS, IRAP, Toulouse, France
   \and
   Institut des Sciences Moléculaires (ISM), CNRS, Univ. Bordeaux, 351 cours de la Libération, 33400 Talence, France
   \and
   Laboratoire d’astrophysique de Bordeaux, Univ. Bordeaux, CNRS, B18N, allée Geoffroy Saint-Hilaire, 33615 Pessac, France
   \\\email{antoine.espagnet@gmail.com}}
   \date{Received 18 March 2026 / Accepted 16 October 2026}
 
  \abstract
   {Some protostars are known to undergo episodic increases in accretion. The resulting luminosity bursts and outbursts heat the inner regions of their envelopes. Previous theoretical and observational studies have shown that the increase in temperature alters the position of the snowlines, which return to their original locations on different timescales depending on the species. However, the chemical effects of luminosity outbursts in protostellar envelopes, including their long-term consequences, remain poorly explored.}
   {This study aims to characterize the chemical impact of a luminosity outburst on the envelope of a solar-type protostar, considering the dynamics of the system, and identify trends in the molecular response relative to both ongoing and past outburst activity.}
   {We used the dynamic semi-analytic code \texttt{Analytical Protostellar Environment (APE)} in combination with the three-phase chemical model NAUTILUS to characterize the time-dependent evolution of molecular abundances in a collapsing envelope, from 10 to $10^4$ au. The luminosity outburst was parameterized by increasing the protostellar luminosity by a factor of 100 lasting 100 years. We predicted the time evolution of the abundances of 83 molecules.}
   {The outburst pushes back molecular snowlines by up to a factor of ten and induces widespread gas phase enrichment through thermal desorption. All species return to their pre-outburst state within a few thousand years at most for $r \leq 1000$ au and remain in that state thereafter. However, eight molecules (\ce{HOCH2CHO}, \ce{CH3COCH2OH}, \ce{C2H5CHO}, OCS, \ce{SO2}, \ce{C2S}, \ce{C3S}, and \ce{HC3N}) exhibit abundance deviations 10–50 kyr after the outburst in the warm inner regions ($T\gtrsim$ 100 K), brought by the inward transport of chemically altered material. Eight species (\ce{CH3O}, \ce{C2O}, \ce{NH2}, \ce{HCCO}, \ce{HCNO}, \ce{HNO}, \ce{NO}, and \ce{NCO}) exhibit an increase in their gas phase abundances in an intermediate region located between $\sim$300 and $\sim$1500 au.}
   {Most of the molecules considered in this study are only impacted by the outburst through an outward shift of their snowlines. A few molecules present other responses to the outburst such as a modification of their abundance in the warm inner region of the envelope a few tens of thousand years after the outburst. With the exception of \ce{HOCH2CHO}, \ce{CH3COCH2OH}, and \ce{C2H5CHO}, all complex organic molecules (COMs) are expected to exhibit similar abundances in the warm inner regions of protostars, regardless of whether they undergo outbursts.}


   \maketitle
   \nolinenumbers

\section{Introduction}

The chemical evolution of the gas and icy mantles of dust grains during the formation of stars plays a key role in shaping the composition of the material from which planets emerge. As the physical conditions within the collapsing system evolve, its chemistry responds accordingly, driven by variations in temperature, density, and radiation field. Observations in the radio, millimeter, and submillimeter wavelengths provide valuable constraints on the molecular content of star-forming regions \citep[e.g.,][]{Jorgensen2016,Marchand2024}. These observations show that the chemical contents of protostars may vary, which could be due to their environments or physical histories \citep[e.g.,][]{Yang2021,Hsu2022}, as suggested by chemical models \citep[e.g.,][]{Coutens2020,Marchand2025b}. The cost of obtaining a detailed chemical inventory of an individual young stellar object (YSO) has prevented the conduct of statistical studies across different evolutionary stages and environments, which are keys to understanding how the aforementioned factors can influence the chemistry. While there are previous and ongoing projects with the NOrthern Extended Millimeter Array (NOEMA) and the Atacama Large Millimeter/submillimeter Array (ALMA), such as COMPASS\footnote{\url{https://erda.ku.dk/vgrid/COMPASS/}} \citep{Jorgensen2026}, modeling the chemical evolution is also an important tool. Consequently, it is particularly important to model phenomena that can occur during star formation in order to understand the observed chemical discrepancies they trigger and to infer the physical history of protostellar systems.
\begin{figure*}
\parbox{5cm}{\caption{Schematic illustration of the outward shift of snowlines during a luminosity outburst. Black shapes and white edges represent, respectively, dust grains and their icy mantles. Left: Quiescent phase (i.e., without a luminosity outburst). Right: Outburst phase. The rectangular box illustrates the thermal desorption process.}\label{fig:impact_burst}}
\parbox{12cm}{\hspace{0.5cm}\includegraphics[width=12cm]{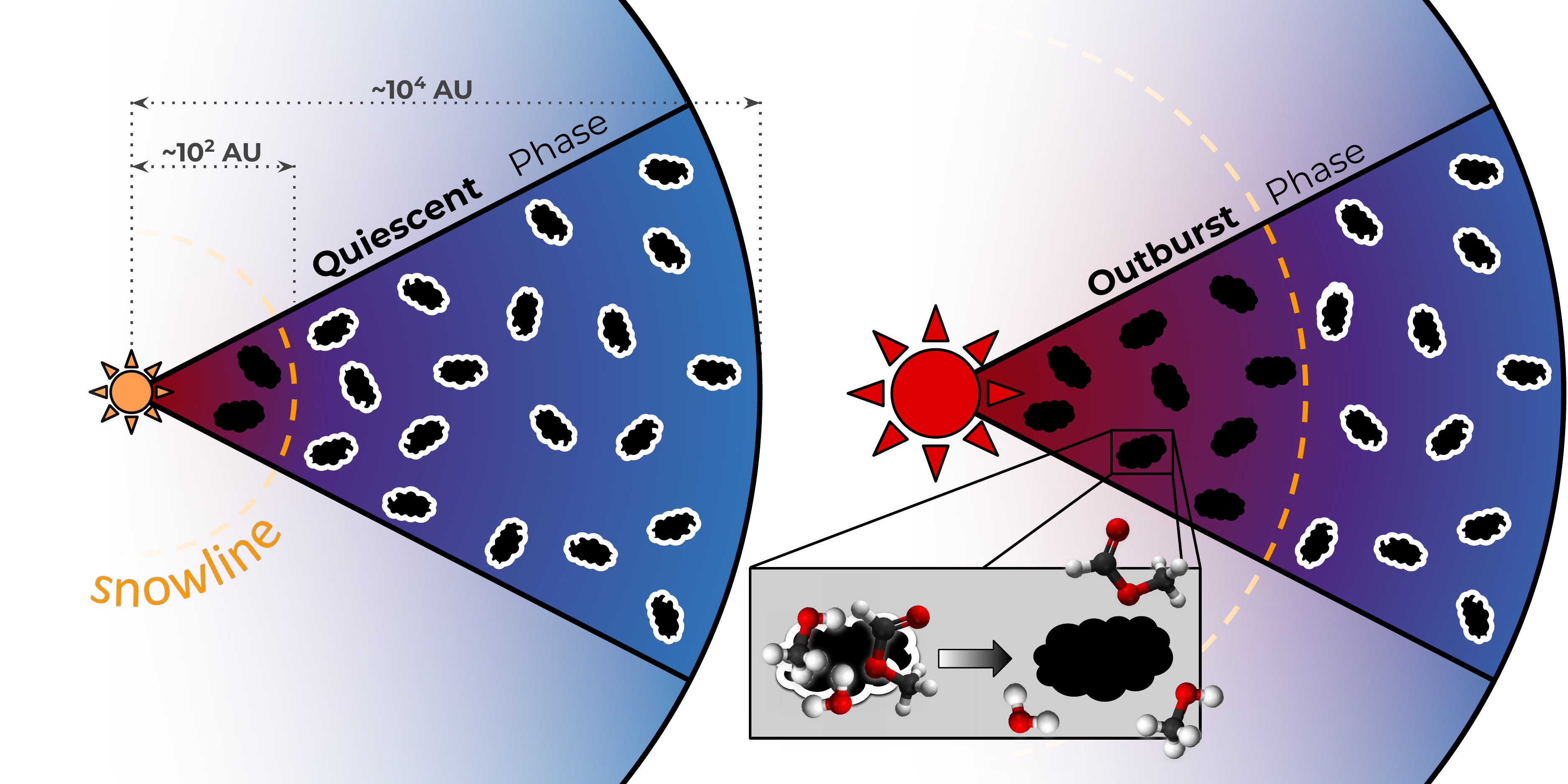}}
\end{figure*}\\
Episodic accretion is a process that can take place during the protostellar evolution. When it occurs, YSOs undergo luminosity bursts or outbursts caused by short-term increases in accretion onto the central protostar. FUor and EXor objects are the most prominent examples of such events \citep{Audard2014, Fischer2023}. The luminosity during those outbursts can increase by a factor of a few tens to a couple of hundreds, with typical durations of a few decades up to a century. We refer to Table 1 of \cite{Fischer2023} for further details on the definition, classification, and differences between bursts, outbursts, and FUor and EXor objects. Episodic accretion explains the wide range of observed low luminosities protostars and how they succeed to accrete their mass during their evolution \citep[e.g.,][]{Evans2009, Audard2014, Fischer2023}. One of the key questions regarding these events is how they shape the physicochemical evolution of protostellar envelopes.\\
The warm inner regions of low-mass protostars are particularly rich in complex organic molecules (COMs), carbon-bearing molecules composed of at least six atoms \citep[e.g.,][]{Herbst2009}, and are called hot corinos. These molecules, considered as precursors to prebiotic chemistry, are believed to form efficiently on icy grain mantles and to sublimate where the dust temperature exceeds their desorption temperature of $\sim$100 K \citep[e.g.,][]{Garrod2007}. For solar-type protostars, this region lies within a few tens to hundreds of astronomical units (au) from the protostar \citep{Ceccarelli2004}.\\
During a luminosity outburst, the increase in luminosity heats the envelope, moving the snowlines (i.e., the interface between the gas and solid phases for a given species) outward (see Fig. \ref{fig:impact_burst}). Once the outburst finishes, temperatures decrease and molecules freeze again on dust grains within several hundred years to a few tens of thousands of years, depending of their location in the envelope (see Sect. \ref{subsec:inner_spatial_variations}). To date, it remains unclear whether the temporary release of molecules into the gas phase, combined with the dynamics, has a long-term impact on both the gas and ice mantle compositions. The associated temperature increase may also alter the chemical composition by enhancing or reducing specific reaction rates, leading to a different evolution of the chemical reservoirs.\\
Several studies have modeled the chemical impact of accretion outbursts on protostellar environments. \citet{Visser2015} investigated the response of simple molecules such as CO, \ce{HCO+}, and \ce{N2H+} to luminosity outbursts, showing that snowlines can be expanded by an order of magnitude, leading to strong abundance variations. \citet{Taquet2016} extended models to a few COMs, demonstrating that past outbursts can induce the formation of some COMs in extended regions of protostellar envelopes. More recently, \citet{Molyarova2018} and \citet{Zwicky2024} explored chemical tracers of outbursts in Class I/II disks, identifying molecules such as \ce{H2CO}, \ce{HCN}, \ce{CO}, \ce{N2H+}, \ce{HCO+}, and \ce{NH2OH} (although the latter has not been detected in disks, \citealt{Ligterink2018}) that can potentially be used to probe past outbursts. However, the chemistry of the envelope of low-mass protostars during and after luminosity outbursts remains poorly explored, particularly across a wide range of observed species.\\
In this work, we investigate the impact of luminosity outbursts on the chemical structure of a solar-type protostar from the warm inner region to the cold outer envelope during the Class 0 and Class I stages. We consider a wider set of molecules, including COMs that can be compared with existing and future interferometric observations. Our goal is to better understand the chemical changes that the outbursts induce and to identify chemical features characteristic of luminosity outbursts.\\
In Sect. \ref{sec:model}, we describe our physicochemical model. Section \ref{sec:results} presents the predicted chemical response to luminosity outbursts. In Sect. \ref{sec:discussion}, we compare our results with previous studies, including observations, and discuss our model limitations. Finally, we summarize our conclusions in Sect. \ref{sec:conclusion}.

\section{Model} \label{sec:model}

In this section, we describe the physicochemical model employed to investigate the chemical impact of luminosity outbursts. We consider different particle trajectories within a collapsing protostellar envelope. These trajectories are chosen to sample the envelope from 10 au to $10^{4}$ au. In the innermost region (10 au), the temperature reaches $\sim250-300$ K (see Fig. \ref{fig:T_r_n_r_APE}), while in the outermost region ($10^{4}$ au), the temperature is 10 K. The chemical model then receives as an input the time evolution of the physical properties (density, temperature, and visual extinction) of the particles, as determined by the physical model. The chemical evolution of particles is calculated for a case without any luminosity outburst (quiescent model) and for nine cases with a luminosity outburst (outburst models) occurring at different times. A sketch summarizing how the model works is presented in Fig. \ref{fig:diagram}.

\subsection{Physical model} \label{subsec:physical_model}
We consider two types of physical models: the quiescent models, without any luminosity outburst (Sect. \ref{subsubsec:quiescent_model}), and the outburst models, exhibiting one luminosity outburst (Sect. \ref{subsubsec:outburst_impacted_model}).

\subsubsection{Quiescent models}
\label{subsubsec:quiescent_model}
\begin{figure}
    \centering
    \includegraphics[width=8.5cm]{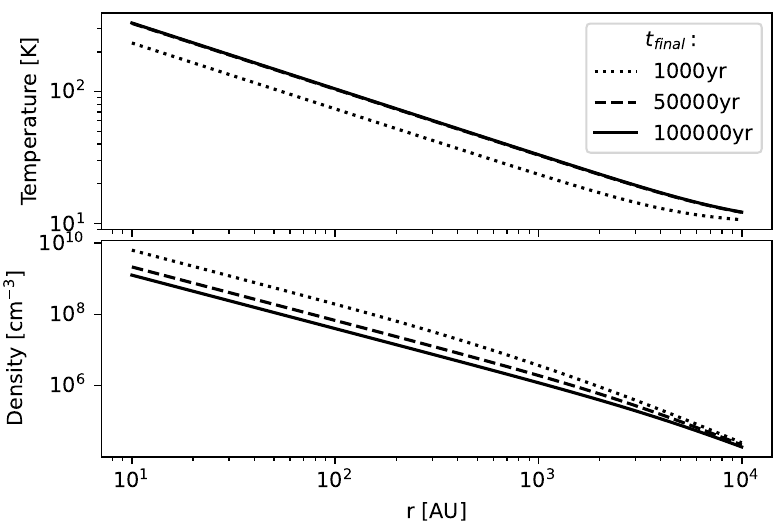}
    \caption{Radial profiles of temperatures and densities of the model. Both panels (top for temperature and bottom for density) show three different $t_\mathrm{final}$ times, which mark the age of the protostar (see Section \ref{subsubsec:quiescent_model}). In the top panel, the 50 kyr and 100 kyr epochs overlap.}
    \label{fig:T_r_n_r_APE}
\end{figure}

The physical model was produced with the \texttt{Analytical Protostellar Environment (APE\footnote{\url{https://bitbucket.org/pmarchan/ape_code}})} code \citep{Marchand2025a}. \texttt{APE} is a dynamic semi-analytic physical model that simulates physical properties of a protostar\footnote{We use the term protostar to designate the central object of the protostellar system. The protostellar system then refers to the entire system, including the protostar, the envelope, the disk, and the jets.} and the surrounding envelope, disk, and outflow. For our study, we employed the \texttt{Particle mode} of \texttt{APE} \citep[described in Sect.\ 2.2.1 in][]{Marchand2025a} in which a particle is falling in a collapsing, and hence evolving, system. This mode tracks the evolution of the density, the temperature, and the visual extinction of a particle during the collapse up to a specified location and time within the protostellar system. By setting the positions of the particles at the final time of the simulation, we retrace their trajectories backward in time (“reverse” option of APE, see \citealt{Marchand2025a}). We chose 20 log-spaced final positions (see $r_\mathrm{final}$ in Table \ref{tab:parameters}), from 10 to 10,000 au. We considered a 1D description by setting the angle between the disk mid-plane and the particles, $\alpha$, to 45° as we wanted to study exclusively the envelope. In this way, we avoided particles that end in the protoplanetary disk or in the outflow. Parameters and values used in this study are detailed in Table \ref{tab:parameters}.\\
The collapse model of APE assumes a free-falling Bonnor-Ebert Sphere \citep{Ebert1955,Bonnor1956}. As the protostellar system is embedded in a molecular cloud, we considered two radiative components heating the particle: the protostar and a background radiation, coming from the surrounding molecular cloud. The temperature at a given distance, $r$, from the protostar is defined by
\begin{equation}
T_\mathrm{quiescent}(r) = \left[ \left(\frac{R_\mathrm{ps}}{r}\right)^{2} T_\mathrm{ps}^4 + T_\mathrm{mc}^4 \right]^{\frac{1}{4}}
\label{eq:quiescent_temperature}
,\end{equation}
where $R_\mathrm{ps}$ and $T_\mathrm{ps}$ are, respectively, the radius and the surface temperature of the protostar. $T_\mathrm{mc}$ is the temperature of the surrounding molecular cloud and is equal to 10 K. A demonstration of this equation is presented in Appendix \ref{app:eq1&2}. As explained in \citet{Marchand2025a}, the temperature was computed using the Stefan-Boltzmann law assuming that the envelope is optically thin to the infrared emission of the protostar. This approach provides a good approximation to the envelope temperature calculated by the radiative transfer computations of RADMC-3D \citep{Marchand2025a}. Furthermore, it is assumed that the gas and dust grains reach thermal equilibrium quickly \citep{Goldsmith2001}, which means that their temperatures are equal (see Sect. \ref{subsec:model_limitations}). We refer to \cite{Marchand2025a} for a detailed description of the model. Figure \ref{fig:T_r_n_r_APE} shows the radial profiles of the temperature and the density for our quiescent model at different times.\\
By default, the envelope in APE dissipates after two free-fall times, as its outer boundary reaches the center. Here, we increase the envelope lifetime by extending its initial radius to $\sim23,000$ au, larger than the default model by a factor of $\sim 2.3$\footnote{We extended the initial size of the envelope in order to increase its lifetime so its outer boudary is still larger than 10,000 au at the final time of the models (up to 150 kyr after the birth of the protostar). In APE, we took the minimal integer value of the normalized radius of the initial Bonnor-Ebert sphere, $x_{max}$, that satisfies such an extension: $x_{max}$=15.}. This ensures that the envelope is present at all points and times considered in this study, i.e., that the envelope still reaches 10,000 au at 150,000 yr after the formation of the outburst.

\begin{figure}
    \centering
    \includegraphics[width=8cm]{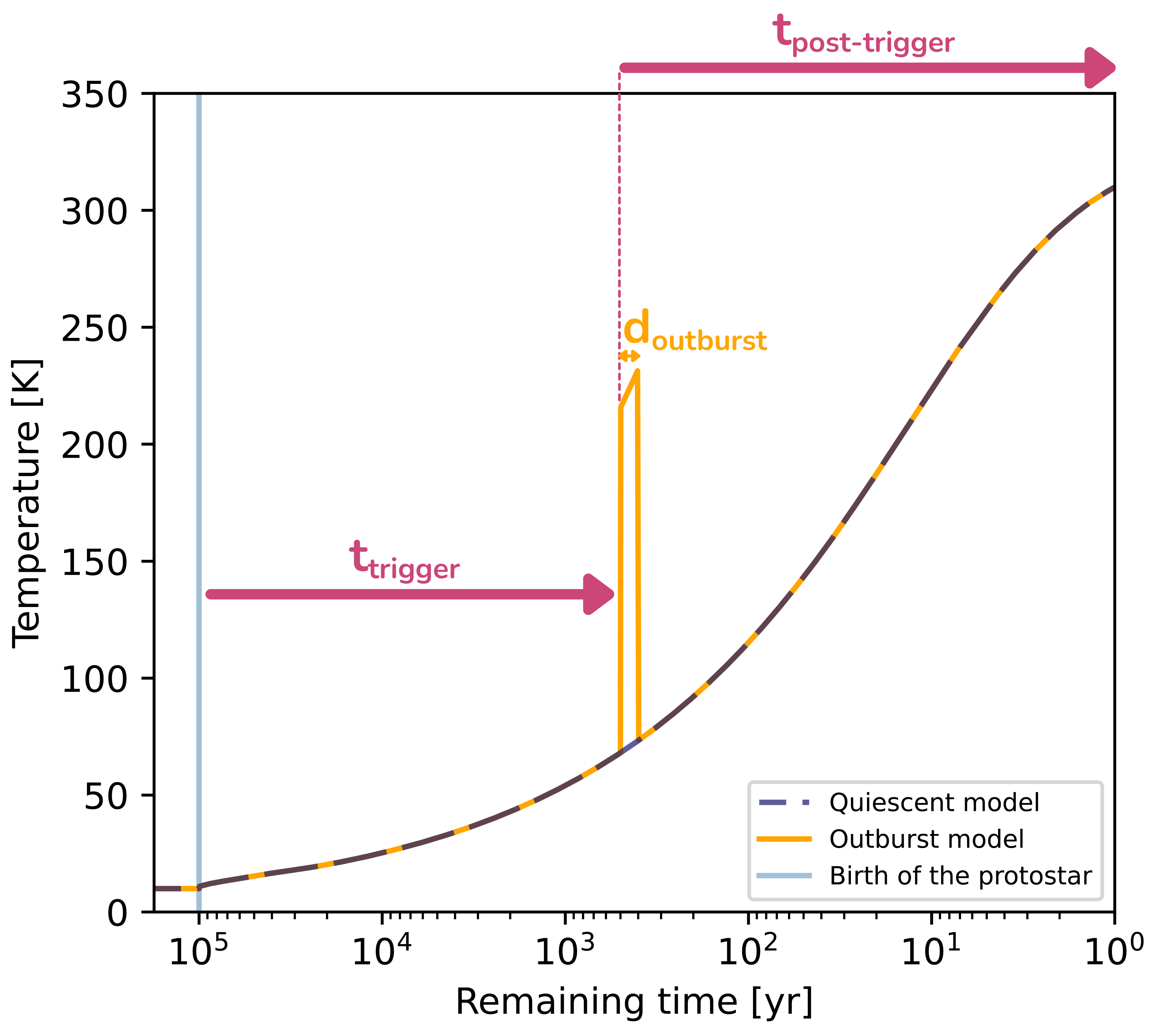}
    \caption{Time evolution of the temperature of a particle reaching $r_\mathrm{final} = 10\ \mathrm{au}$ at the end of the simulation. The semitransparent dashed black line represents the quiescent model and the solid orange line is for the outburst model. The birth of the protostar is represented by the vertical blue line. Values used for $t_{\mathrm{trigger}}$, $t_{\mathrm{post-trigger}}$, and $d_{\mathrm{outburst}}$ are presented in Table \ref{tab:parameters}. The x axis shows the time remaining until the end of the simulation, providing a clear view of the later stages of the simulation, during which the temperature evolves rapidly.}
    \label{fig:Physical_model_quiescent_outburst}
\end{figure}
\begin{table*}
  \centering
  \begin{threeparttable}[b]
   \caption{Main parameters of the model.\protect\footnotemark}
    \begin{tabular}{lll}
        \hline
        \hline
        \textbf{Parameter} & \textbf{Description} & \textbf{Value}\\
        \hline
          &\textbf{Physics} &\\
        \hline
        $M_\mathrm{{BE}}$ &Mass of the Bonnor-Ebert sphere& $1\ \mathrm{M_\odot}$\\
        $A_\mathrm{V\ mc}$ &Visual extinction of the surrounding molecular cloud&$5\ \mathrm{mag}$\\
        $\alpha$&Angle between the disk mid-plane and the particle& $45\ \mathrm{^{\circ}}$\\
        $T_\mathrm{mc}$&Temperature of the sourrounding molecular cloud& $10\ \mathrm{K}$\\
        $r_\mathrm{final}^*$&Final distance of the particle from the protostar (20 log-spaced values)&$10-10000\ \mathrm{au}$\\
        $t_\mathrm{trigger}^*$ & Triggering time of the outburst relative to the birth of the protostar& $10^3-10^5\ \mathrm{yr}$\\
        &(10 lin-spaced values)&\\
        $t_\mathrm{post-trigger}^*$ & Final time of the simulation relative to the trigger of the outburst& $10,\ 100,\ 200,\ 500\ \mathrm{yr, and }\ 1,$\\
         &(time after the beginning of the outburst)& $5,\ 10,\ 20,\ \mathrm{and}\ 50\ \mathrm{kyr}$\\
         & Additional models for $r_\mathrm{final} =\ $10 au & $15,\ 25,\ 30,\ 35,\ 40,\ \mathrm{and}\ 45\ \mathrm{kyr}$\\
        $f_\mathrm{L}$&Ratio of the protostellar luminosity between the outburst and quiescent phase&$100$\\
        $d_\mathrm{{outburst}}$&Duration of the luminosity outburst&$100\ \mathrm{yr}$\\
        $d_\mathrm{{increase}}$&Duration of the increase in luminosity until it reaches its maximum value&$1\ \mathrm{yr}$\\
        $d_\mathrm{decrease}$&Duration of the decrease in luminosity once the outburst has ended&$1\ \mathrm{yr}$\\
        \hline
          &\textbf{Chemistry}&\\
        \hline
        $D/M$&Dust-to-gas mass ratio&$1\ \mathrm{\%}$\\
        $r_\mathrm{g}$&Grain radius&$0.1\ \mathrm{\mu m}$\\
        $\rho_\mathrm{{g}}$&Mass density of grain material&$3\ \mathrm{g}~\mathrm{cm}^{-3}$\\
        $k_\mathrm{CR}$&Cosmic ray ionisation rate&$1.3\times10^{-17}\ \mathrm{s^{-1}}$\\
        $S_\mathrm{n}$&Sticking coefficient for neutral species&$1$\\
        $S_\mathrm{i}$&Sticking coefficient for ions &$0$\\
        \hline
    \end{tabular}
    \label{tab:parameters}
  \end{threeparttable}
\end{table*}

\subsubsection{Outburst models}
\label{subsubsec:outburst_impacted_model}
\begin{figure}
    \centering
    \includegraphics[width=8.5cm]{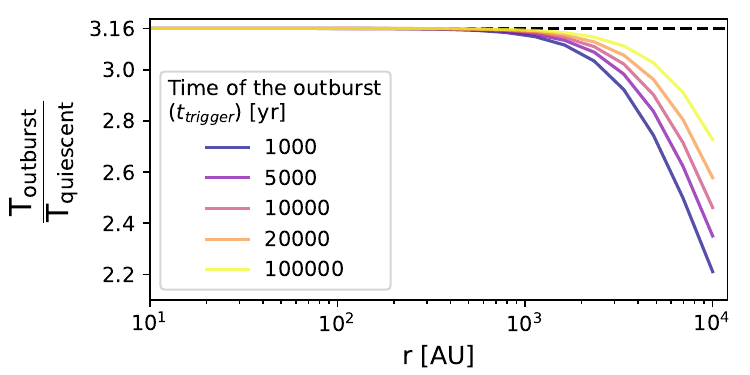}
    \caption{Temperature ratios between the outburst and quiescent models, evaluated for different triggering times of the outburst, as a function of radius.}
    \label{fig:T_burst_T_quiescent_profile}
\end{figure}

To simulate a luminosity outburst, we increased the protostar luminosity by a factor of 100 for 100 years, which correspond to typical values of a FUor outburst \citep{Fischer2023}. These values are also used in other models \citep[e.g.,][]{Johnstone2013,Visser2015,Taquet2016,Zwicky2024}. We added a linear increase and decrease in the temperature, lasting 1 year, respectively at the beginning and at the end of the outburst to transition from the quiescent to the outburst state and back. The time at which the outburst occurs during the simulation is defined by the user. Consequently, we defined two distinct time intervals, $t_\mathrm{trigger}$ and $t_\mathrm{post-trigger}$, corresponding, respectively, to the period from the formation of the protostar to the onset of the outburst, and from the onset of the outburst to the end of the simulation. Thus $t_{\mathrm{final}}$ is the sum of $t_{\mathrm{trigger}}$ and $t_{\mathrm{post-trigger}}$. These parameters are presented in Figure \ref{fig:Physical_model_quiescent_outburst}, which displays the temperature evolution of the quiescent and outburst models as a function of time for a representative particle. In our physical model, the luminosity outburst only impacts the temperature. The temperature during the luminosity outburst was calculated as
\begin{equation}
T_\mathrm{outburst}(r) = \left[ f_\mathrm{L} \left(\frac{R_\mathrm{ps}}{r}\right)^{2} T_\mathrm{ps}^4 + T_\mathrm{mc}^4 \right]^{\frac{1}{4}}
\label{eq:outburst_temperature}
,\end{equation}
with the luminosity increase factor $f_\mathrm{L}$ = 100 (see Appendix \ref{app:eq1&2}). Therefore the closer a particle is to the protostar, the more its temperature is impacted by the outburst. Furthermore, the protostellar surface temperature and radius evolve with its age (see Fig. \ref{fig:T_r_n_r_APE}). As a result, both outburst and quiescent temperatures vary with the protostellar age, leading to a time evolution of the $T_\mathrm{outburst}/T_\mathrm{quiescent}$ ratio. Figure \ref{fig:T_burst_T_quiescent_profile} shows the ratio between the outburst and the quiescent temperature as a function of radius for five different times. In the warm inner part of the envelope, the outburst increases the temperature by a factor of $\sim f_\mathrm{L}^{1/4} \approx 3.16$, and in the cold outer envelope by a factor of $\sim 2.2-2.7$ at $10{,}000$ au.

\subsection{Chemical model} \label{subsec:chemical_model}
We used the Nautilus three-phase gas-grain chemical code \citep{Ruaud2016}, which computes the time evolution of chemical abundances according to the evolution of physical parameters (temperature, $T$, density, $n$, and visual extinction, $A_\mathrm{V}$) of each particle. It considers the gas phase and two chemically active phases for the dust grains: the mantle and the surface, with possible exchanges allowed between those phases. The chemical network used in this study contains $\sim$800 species and $\sim$9,000 reactions \citep{Manigand2021,Coutens2022,Agundez2023,Marchand2025a}. Desorption can proceed via both thermal and nonthermal mechanisms, including cosmic-ray and UV-induced processes, chemical desorption, and ice sputtering resulting from cosmic-ray collisions \citep{Wakelam2021}. The visual extinction, $A_\mathrm{V}$, parametrizes the attenuation of the UV interstellar radiations by the presence of gas and dust along the line of sight, and therefore regulates the local photodissociation, photoionization, and photodesorption rates in the model. It is composed of two components: a constant component from the surrounding molecular cloud, $A_\mathrm{V\ mc}$ = 5 mag, and a variable component depending on the position of the particle in the protostellar system \citep[Sect. 2.7]{Marchand2025a}.\\
As presented in Table \ref{tab:parameters}, we assume a sticking probability of 1 for neutral species and of 0 for ions. The diffusion energies are set to be a fraction of the binding energy for the surface (0.4) and the mantle (0.8). All the species can diffuse by the tunnel effect depending on their mass \citep{Wakelam2024}, with a diffusion barrier thickness of 0.25 nm. We assume spherical grains with a constant radius set to ${r_\mathrm{grain}} = 0.1\mathrm{~\mu m}$, a bulk density of $\rho_\mathrm{grain} = 3\mathrm{~g\ cm^{-3}}$, and a dust-to-gas mass ratio of 1$\%$.\footnotetext{The varied parameters are indicated with $^*$.}\\
Before running the chemistry on quiescent and outburst models (see Sect. \ref{subsubsec:quiescent_model} and \ref{subsubsec:outburst_impacted_model}), we ran a physically static simulation corresponding to the chemical evolution occurring in a molecular cloud for a duration of 1 million years. The initial abundances and physical parameters used for this static simulation are listed in Table \ref{tab:static_parameters}. The abundances are similar to those used in \cite{Hincelin2011}, with differences for S, Si, and P. This was done to better model sulfur compounds by minimizing the distance of disagreement, as defined in \citealt{Wakelam2006}, and to reproduce the upper limits for SiO \citep{Ziurys1989} and PN \citep{Turner1990} in a molecular cloud such as TMC-1. The initial abundances of N and N$_2$ were set to limit the production of nitriles in protostars for better agreement with observations. These changes did not affect oxygenated COMs. The resulting chemical abundances were then used as initial abundances for the collapse phase.

\begin{table}[h]
    \centering
    \caption{Physical parameters and initial chemical abundances used for the initial static phase.}
    \begin{tabular}{lllll}
        \hline
        \hline
        \multicolumn{3}{l}{\textbf{Physical parameters}} & \multicolumn{2}{l}{Value} \\
        \hline
        \multicolumn{3}{l}{Temperature} & \multicolumn{2}{l}{$10\ \mathrm{K}$} \\
        \multicolumn{3}{l}{Density} & \multicolumn{2}{l}{$2\times10^4\ \mathrm{cm^{-3}}$} \\
        \multicolumn{3}{l}{Visual extinction} & \multicolumn{2}{l}{$5\ \mathrm{mag}$} \\
        \multicolumn{3}{l}{Duration} & \multicolumn{2}{l}{$10^6\ \mathrm{yr}$} \\
        \hline
        \multicolumn{3}{l}{\textbf{Initial abundances}} & \multicolumn{2}{l}{(relative to H)} \\
        \hline
        He & $9\times10^{-2}$ & & N & $2\times10^{-6}$ \\
        $\mathrm{N_2}$ & $9\times10^{-5}$ & & O & $2.4\times10^{-4}$ \\
        $\mathrm{H_2}$ & $0.5$ & & C$^+$ & $1.7\times10^{-4}$ \\
        S$^+$ & $2\times10^{-7}$ & & Si$^+$ & $6\times10^{-11}$ \\
        Fe$^+$ & $2\times10^{-8}$ & & Na$^+$ & $2\times10^{-8}$ \\
        P$^+$ & $1\times10^{-8}$ & & Cl$^+$ & $1\times10^{-7}$ \\
        F & $6.7\times10^{-9}$ & & & \\
        \hline
    \end{tabular}
    \label{tab:static_parameters}
\end{table}

\subsection{Grid of models}  \label{subsec:grid_of_models}
\begin{figure}
    \centering
    \includegraphics[width=8.8cm]{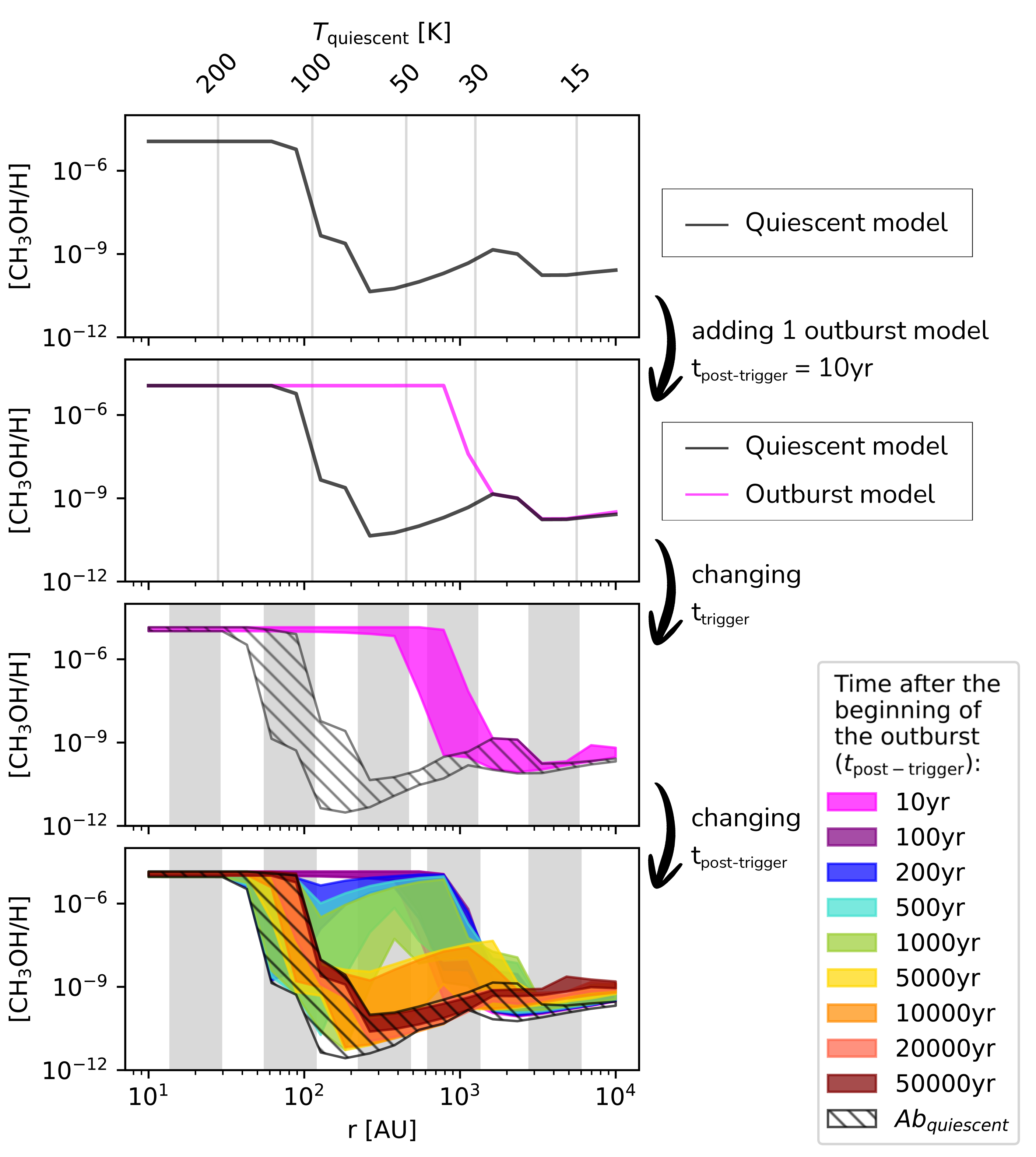}
    \caption{Radial profiles of the gas phase abundance of methanol. The four panels present step by step how we built the grid of models. The different steps are explained in Sect. \ref{subsec:grid_of_models}. The bottom x axis indicates the distance between the particle and the protostar. The top x axis indicates the corresponding quiescent temperature. The spread of the quiescent temperatures due to the evolution of the protostar is represented through the vertical gray bands. Abundances inherited from the quiescent model are shown in hatched black, while abundances predicted by the outburst model are shown in color. The different colors indicate the elapsed time since the beginning of the outburst. Note that the pink (10 years) and the purple (100 years) profiles are during the outburst, while the other colors are post-outburst profiles.}
    \label{fig:figure_grid_CH3OH}
\end{figure}

\begin{figure*}
    \centering
    \includegraphics[width=1\textwidth]{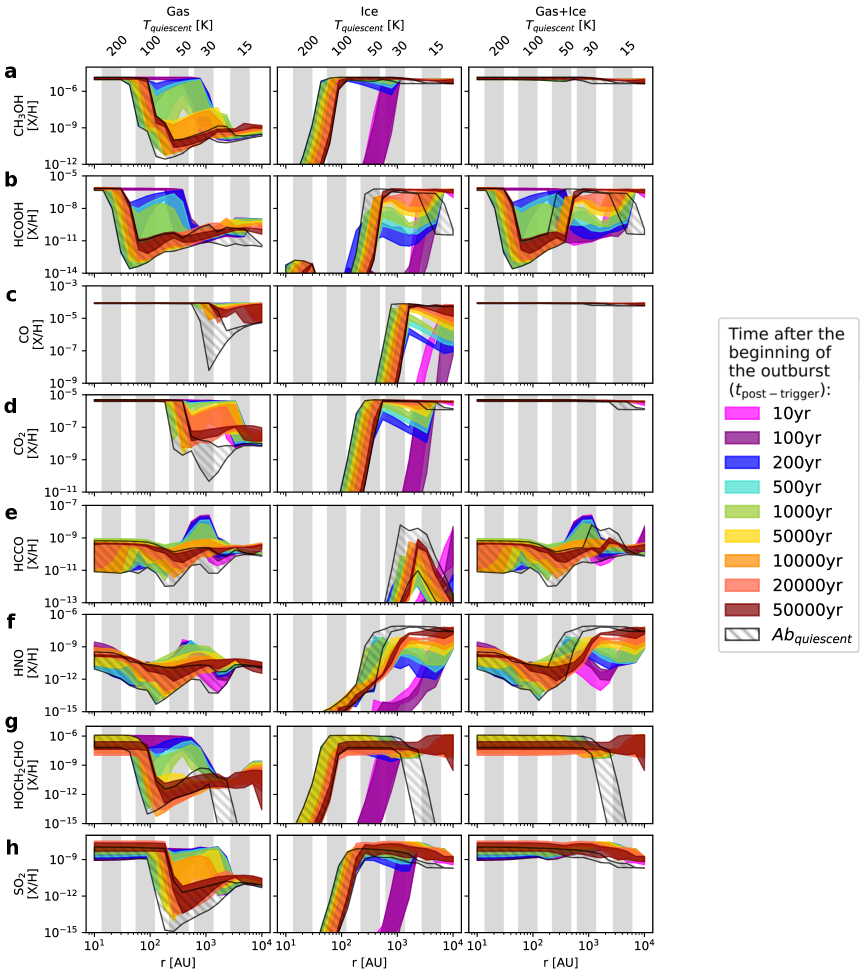}
    \caption{Radial abundance profiles of (a) \ce{CH3OH}, (b) \ce{HCOOH}, (c) \ce{CO}, (d) \ce{CH4}, (e) \ce{HCCO}, (f) \ce{HNO}, (g) \ce{HOCH2CHO}, and (h) \ce{SO2}. Gas, ice, and total abundances in the left, middle, and right columns, respectively. The bottom x axis indicates the distance between the particle and the protostar at the final time of the simulation. The top x axis indicates the corresponding quiescent temperature. The spread of the quiescent temperatures due to the evolution of the protostar is represented through the vertical gray bands (see details in Sect. \ref{sec:model}). Abundances of the quiescent models are shown in hatched black, while abundances impacted by a luminosity outburst are shown in color.}
    \label{fig:radial_profile}
\end{figure*}

In addition to considering different final positions $r_\mathrm{final}$ (see Sect. \ref{subsubsec:quiescent_model}), we also varied the time at which the outburst was triggered ($t_{\mathrm{trigger}}$) and the final time of the simulations (relative to the trigger of the outburst, $t_{\mathrm{post-trigger}}$). As $t_{\mathrm{final}} = t_{\mathrm{trigger}} + t_{\mathrm{post-trigger}}$ (see Fig. \ref{fig:Physical_model_quiescent_outburst}), $t_{\mathrm{final}}$ also varies. The values of the triggering time of the outburst, $t_{\mathrm{trigger}}$, (see Table \ref{tab:parameters}) were chosen to study different stages in the evolution of the protostellar system, while those of the final time of the simulation relative to the trigger of the outburst, $t_{\mathrm{post-trigger}}$, were selected to investigate the evolution of abundance during and after the outburst. As the outburst is lasting 100 years (see Sect. \ref{subsubsec:outburst_impacted_model}), for the two first tpost-trigger (10 and 100 years), the simulation ends while the luminosity outburst is still ongoing. For simulations where $t_{\mathrm{post-trigger}}$ = 100 yr, the simulation ends at the end of the outburst, before the protostar returns to its quiescent state. Adopting this broad grid of 1800 models minimizes the possible biases that the choice of time parameters might have on the chemical abundances.\\
Figure \ref{fig:figure_grid_CH3OH} shows how we build the grid of models in four steps using the methanol ($\mathrm{CH_3OH}$) molecule as example. The top panel displays the radial profile of gas phase abundance for the quiescent model, i.e., without any luminosity outburst. In the second panel, we add one luminosity outburst model with $t_{\mathrm{post-trigger}}$ = 10 yr. Thus the pink line is the radial profile of the gas phase abundance 10 years after the outburst started. In the third panel we vary the age of the protostar at the onset of the outburst, $t_{\mathrm{trigger}}$, from $10^3$ to $10^5$ years. In this panel, $t_{\mathrm{post-trigger}}$ is still equal to 10 yr. This step introduces a spread in the abundance profiles reflecting the time evolution of the protostellar system. Additionally, the vertical gray lines indicating quiescent temperatures are horizontally spread due to the increase in the temperature profile through time (see Figure \ref{fig:T_r_n_r_APE}). In outburst cases, the temperature is $\sim$2.2 to 3.16 times the $T_\mathrm{quiescent}$ values (see Figure \ref{fig:T_burst_T_quiescent_profile}). Finally, the last panel shows the profiles for the eight remaining $t_{\mathrm{post-trigger}}$ values (100, 200, 500, 1000, 5000, 10000, 20000, and 50000 yr). Each $t_{\mathrm{post-trigger}}$ value is represented by a distinct color in Figs. \ref{fig:figure_grid_CH3OH}, \ref{fig:radial_profile}, \ref{fig:inner_impact}, \ref{fig:inner_impact_ratio}, and \ref{fig:N2H+_HCO+}. It is worth noting that, for most of the species, variations in $t_{\mathrm{post-trigger}}$ have a bigger impact on the predicted abundances than variations in $t_{\mathrm{trigger}}$, as can be seen for methanol in the last panel of Fig. \ref{fig:figure_grid_CH3OH}.

\section{Results}
\label{sec:results}
In this section, we present the various impacts induced by the luminosity outburst. We classify the types of impacts into four different categories. In the first one, we examine the spatial variations in the snowlines induced by thermal desorption. In the second, we investigate a comparable type of spatial variation that is produced by other chemical processes. In the third one, we present a gas-phase abundance excess observed for eight molecules between $\sim$300 and $\sim$1500 au in response to the outburst. Finally, for the fourth category, we detail the impacts we observe on the abundances more than 10 kyr after the outburst. Our analysis focuses on a set of 83 molecules, which are selected based on two criteria: their detection toward protostars and their predicted abundances (at least $10^{-12}$ relative to hydrogen). The radial abundance profiles of 12 molecules are presented in Fig. \ref{fig:radial_profile} and \ref{fig:N2H+_HCO+}. In addition, the abundance profiles of the other molecules are available on \url{https://zenodo.org/records/23667779}. Table \ref{tab:results_appendix} lists all the species and their corresponding chemical responses to the outburst.

\subsection{Spatial extent of the snowlines}
\label{subsec:inner_spatial_variations}
Most of the molecules (47 out of 83) exhibit an uniform and constant high gas-phase abundance from the inner location of our model, at 10 au, to a location corresponding to the snowline of the considered species. These regions are called “abundance plateaus” in the rest of this paper. Beyond this plateau, the gas-phase abundance decreases significantly. We note that for COMs the plateau region corresponds to the hot corino of the protostellar system. In addition to the abundance plateaus, these molecules exhibit a constant total (ice + gas) abundance profile throughout the whole envelope. Consequently, we also observe an ice-phase abundance plateau from the location of the snowline of the species, to the outer location of our model, at $10^{4}$ au. When the outburst occurs, the gas-phase abundance plateau expands further away in the envelope. Accordingly, the ice-phase abundance plateau is pushed farther out into the envelope.\\
The 47 molecules are the following: \ce{C2H5CHO}, \ce{C2H5OCH3}, \ce{CH3CHO}, \ce{CH3COCH2OH}, \ce{CH3COCH3}, \ce{HC3N}, \ce{HCOOCH3}, \ce{HOCH2CHO}, \ce{NH2CHO}, \ce{OCS}, \ce{SO2}, \ce{C3S}, \ce{CO}, \ce{C2H3CHO}, \ce{C2H5CN}, \ce{C2H5OH}, \ce{C3H6}, \ce{C3O}, \ce{CH2NH}, \ce{CH3CCH}, \ce{CH3CN}, \ce{CH3COOH}, \ce{CH3Cl}, \ce{CH3NC}, \ce{CH3NCO}, \ce{CH3OCH2OH}, \ce{CH3OCH3}, \ce{CH3OH}, \ce{CH3SH}, \ce{CNCN}, \ce{CO2}, \ce{H2CCO}, \ce{H2CO}, \ce{H2CS}, \ce{H2O}, \ce{HOC2H4OH}, \ce{HOCH2CN}, \ce{HOCHCHCHO}, \ce{HONO}, \ce{N2O}, \ce{NH2CN}, \ce{NH3}, \textit{c}-\ce{C2H4O}, \textit{l}-\ce{C3H2}, \ce{CH4}, \ce{NS}, and \textit{c}-\ce{C3H2}. As a representative example of these molecules, row a of Fig. \ref{fig:radial_profile} shows the gas phase, ice-phase, and total abundances of methanol (\ce{CH3OH}). In the quiescent model (hatched black), from the outer to the inner envelope, the ice-phase abundance exhibits a steep decrease starting at around 50–100 au (corresponding to 150–100 K), while the gas phase abundance increases accordingly. Methanol is thus mainly present in the gas phase within the plateau region (the sublimation zone) and locked in the ices beyond it, where the ice abundance exceeds the gas abundance by several orders of magnitude. The total \ce{CH3OH} abundance remains constant throughout the envelope, indicating that the variations between the gas and ice phases result mainly from sublimation. In addition to the verification of the associated reactions, we can rule out the other types of desorption as the main reason for the expansion of the abundance plateau as explained hereafter. In our model, the outburst only affects the temperature profile (see Sect. \ref{subsubsec:outburst_impacted_model}), while nonthermal desorption processes (e.g., UV induced desorption) remain unchanged, since we do not account for any variation in the protostellar UV field (see Sect. \ref{subsec:model_limitations} and Appendix \ref{subsec:UV}) or for negligible dynamical effects such as winds that could modify the density or extinction profiles. Furthermore, chemical desorption would lead to an increase in the total abundance, as it corresponds to the release of newly formed grain-surface products into the gas phase. However, for the 47 molecules presented in this section, no such increase in total abundance is observed during the outburst compared to the quiescent case, further supporting thermal desorption as the dominant mechanism driving the gas–ice abundance variations.\\
The time needed for molecular abundances to return to its quiescent value depends on the density and the temperature of the medium, and therefore on its position in the envelope. This time, $\tau_{fo}$, corresponds to the freeze-out timescale and can be calculated as \citep{Ruaud2016}
\begin{equation}
    \tau_{fo} = \left[n_\mathrm{g}\pi r_\mathrm{g}^2 \varv_{\mathrm{th}} S_\mathrm{n}\right]^{-1}\\
    \mathrm{with}\\
    \varv_{th} = \left(\frac{8\mathrm{k_b}T_\mathrm{k}}{\pi m_\mathrm{s}}\right)^{\frac{1}{2}}
    \label{eq:freeze_out}
,\end{equation}
where $\varv_{\mathrm{th}}$ is the thermal speed, $T_\mathrm{k}$ the kinetic temperature, and $m_\mathrm{s}$ the mass of the molecule. $n_\mathrm{g}$ and $r_\mathrm{g}$ are, respectively, the number density of grains and the grain radius. Finally, $S_\mathrm{n}$ is the sticking coefficient for neutral species on grains. For the 83 molecules considered, this timescale varies between a few hundred years to one thousand years for the inner 100 au and can go up to several tens of thousand years beyond 1000 au.\\
The desorption temperature of each molecule defines the radial position of the outer edge of its inner gas phase abundance plateau, and is thus linked to its freeze-out time. We refer to Table \ref{tab:results_appendix} for the list of molecules presenting this inner abundance plateau with the predicted gas phase abundance and the corresponding radius and temperature at the outer edge of the plateau, as well as its spatial extension during the outburst.\\
CO, \ce{CO2}, \ce{CH4}, \ce{H2CCO}, \ce{H2CO}, \ce{N2O}, and \ce{OCS} exhibit lower desorption temperatures than water and COMs. Consequently, the gas phase abundance plateaus of these molecules extend farther into the outer regions of the envelope (see rows c and d of Fig. \ref{fig:radial_profile} and Table \ref{tab:results_appendix}).\\

\subsection{Spatial extent of an abundance plateau produced by other mechanisms than direct sublimation}
\label{subsubsec:other_processes}
We also find an abundance plateau for six other molecules: \ce{HCOOH}, \ce{CH2OH}, \ce{H2S}, \ce{HNC}, \ce{HCN}, and \ce{HNCO}. The outer limit of this plateau increases when the outburst occurs. However, for these molecules, the total abundance is not constant, indicating that the expansion of the high gas phase abundance region cannot be attributed mainly to direct thermal desorption. We provide a detailed analysis of \ce{HCOOH} as a representative example of this category of molecules and give the chemical reactions related to the other five species afterward. \ce{HCOOH} gas, ice, and total abundance profiles are displayed in the second row of Fig. \ref{fig:radial_profile} (row b). Its total abundance profile in the quiescent case can be divided into three distinct regions: an inner high-abundance region where T$\gtrsim$150 K, a low-abundance region with temperatures between 50 and 150 K, and an outer high-abundance region with T<50 K. As the grains move from the outer envelope into the inner regions, they experience an increase in temperature. In the region with T $\sim50-150$ K, \ce{HCOOH} interacts with \ce{NH3} on the ices to form \ce{NH4COOH}. As a consequence, the \ce{HCOOH} ice and total abundances are low compared to the region below 50 K. With the collapse, the grains continue inward and as they reach a temperature of 150 K. At this temperature, \ce{NH4COOH} sublimates and immediately reforms \ce{HCOOH} and \ce{NH3} in the gas phase (\textit{s}-\footnote{\textit{s}- refers to the ice phase}\ce{NH4COOH} $\rightarrow$ \textit{g}-\footnote{\textit{g}- refers to the gas phase}\ce{NH3} + \textit{g}-\ce{HCOOH}). We thus note that the region rich in \ce{NH4COOH} constrains the location of the high-abundance region of \ce{HCOOH}, which is determined by the temperature. During the outburst, the increase in temperature shifts this region farther out in the envelope. Therefore, the outburst shifts the high-abundance region of \ce{HCOOH} in the same proportion as for direct thermal desorption. The abundance profiles then return to their quiescent state within a few kyr in the gas phase at radii of $\sim$50--500 au (between $t_\mathrm{post-trigger}=1,000$ and $5,000$ yr), and within a few tens of thousand years in the ice phase at radii of $\sim$500--3000 au ($t_\mathrm{post-trigger}\sim20,000$ yr).\\
Similarly to \ce{HCOOH}, \ce{CH2OH}, \ce{H2S}, \ce{HNC}, \ce{HCN}, and \ce{HNCO} exhibit an abundance plateau that is not produced by thermal desorption but by a chemical reformation following a destruction:
\begin{itemize}
    \item For \ce{CH2OH}, the ices are mostly destroyed through \textit{s}-\ce{CH2OH} + \textit{s}-\ce{H2CO} $\rightarrow$ \textit{s}-\ce{CH3OH} + \textit{s}-\ce{HCO}. The gas phase production of \ce{CH2OH} is due to \textit{g}-\ce{OH} + \textit{g}-\ce{H2CCO} $\rightarrow$ \textit{g}-\ce{CH2OH} + \textit{g}-\ce{CO}.
    \item \ce{H2S} is destroyed in the ice phase through \textit{s}-\ce{NH3} + \textit{s}-\ce{H2S} $\rightarrow$ \textit{s}-\ce{NH4SH} and reformed in the gas phase via \textit{s}-\ce{NH4SH} $\rightarrow$ \textit{g}-\ce{NH3} + \textit{g}-\ce{H2S}.
    \item For \ce{HNC} (and HCN), the reactions are \textit{s}-\ce{NH3} + \textit{s}-\ce{HNC} (or \textit{s}-\ce{HCN}) $\rightarrow$ \textit{s}-\ce{NH4CN} and \textit{s}-\ce{NH4CN} $\rightarrow$ \textit{g}-\ce{NH3} + \textit{g}-\ce{HNC} (or \textit{g}-\ce{HCN}).
    \item For \ce{HNCO}, the reactions are \textit{s}-\ce{NH3} + \textit{s}-\ce{HNCO} $\rightarrow$ \textit{s}-\ce{NH4NCO} and \textit{s}-\ce{NH4NCO} $\rightarrow$ \textit{g}-\ce{NH3} + \textit{g}-\ce{HNCO}.
\end{itemize}
\begin{figure*}
    \centering
    \includegraphics[width=0.80\textwidth]{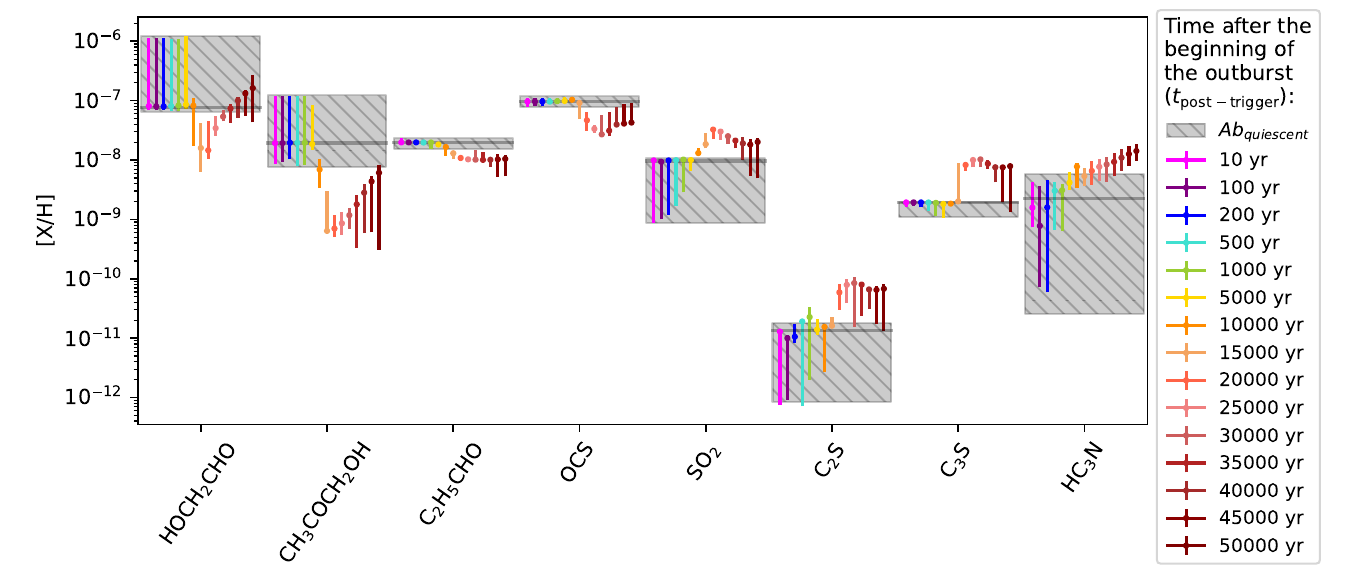}
    \caption{Inner gas phase abundances for the eight molecules that see their abundances impacted by the outburst more than 10 kyr after its onset. The abundances correspond to particles ending at 10 au. The colors indicate different $t_{\mathrm{post-trigger}}$. Dots represent the median gas phase abundances predicted by the models, while vertical lines denote the corresponding minimum and maximum predicted values. The gray rectangles indicate the range of abundances in the quiescent models, with the horizontal black lines marking the median abundance.}
    \label{fig:inner_impact}
\end{figure*}
\begin{figure*}
    \centering
    \includegraphics[width=0.80\textwidth]{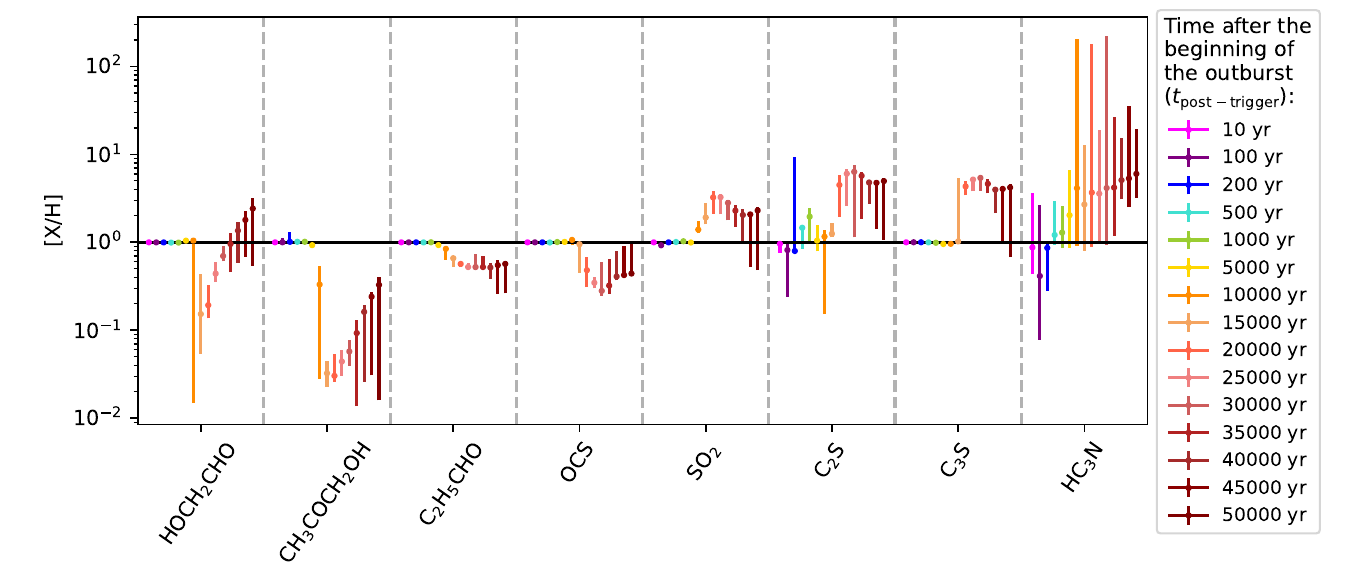}
    \caption{Same as Fig. \ref{fig:inner_impact}, but showing relative abundances instead of absolute values. The y axis represents the abundance ratio of the outburst models relative to the corresponding quiescent models.}
    \label{fig:inner_impact_ratio}
\end{figure*}

\subsection{Gas phase abundance increase between 300 and 1500 au}
\label{subsec:outer_spatial_variations}

In the outburst case, eight species manifest a localized enhancement of gas-phase abundance in an intermediate region between $\sim$ 300 au and $\sim$ 1500 au, corresponding to quiescent temperatures of $\sim$ 30--50 K. This leads to a distinct bump in their radial gas phase abundance profiles. These species are \ce{CH3O}, \ce{C2O}, \ce{NH2}, \ce{HCCO}, \ce{HCNO},  \ce{HNO}, \ce{NO}, and \ce{NCO}. The radial profiles of abundances of two of these molecules (HCCO and HNO) are shown in rows e and f of Fig. \ref{fig:radial_profile}.\\
These molecules are abundant in the ice-phase beyond the region of the bump, except for \ce{NCO}, which is more abundant in the gas phase, and for \ce{HCNO}, for which the ices are not included in our network. In the intermediate region, the abundance of ices decreases and the molecules are then mainly in the gas phase. In the outburst case, the abundance profiles show the emergence of a local gas phase abundance excess in this region. Here, the outburst can change the main reactions that produce and destroy molecules. For example, we find that in the outburst case, \ce{HCCO} is formed more efficiently in the gas phase due to the reaction \textit{g}-\ce{C} + \textit{g}-\ce{H2CO} $\rightarrow$ \textit{g}-\ce{H} + \textit{g}-\ce{HCCO}, causing the increase in abundance that we observe.\\
We present hereafter the corresponding reactions for the other molecules:
\begin{itemize}
    \item For \ce{CH3O}, in the outburst case, the gas-phase excess is produced with \textit{g}-\ce{C2H} + \textit{g}-\ce{CH3OH} $\rightarrow$ \textit{g}-\ce{CH3O} + \textit{g}-\ce{C2H2} and \textit{g}-\ce{CN} + \textit{g}-\ce{CH3OH} $\rightarrow$ \textit{g}-\ce{CH3O} + \textit{g}-\ce{HCN}.
    \item The \ce{C2O} bump is produced by the same reactions that explain the abundance increase at shorter radii, with a higher-efficiency region: \textit{g}-\ce{O} + \textit{g}-\ce{C3O} $\rightarrow$ \textit{g}-\ce{CO} + \textit{g}-\ce{C2O} and \textit{g}-\ce{O} + \textit{g}-\ce{HC3O} $\rightarrow$ \textit{g}-\ce{HCO} + \textit{g}-\ce{C2O}.
    \item This is also the case for the reactions related to the \ce{HNO} excess, with \textit{g}-\ce{NO} + \textit{g}-\ce{HCO} $\rightarrow$ \textit{g}-\ce{HNO} + \textit{g}-\ce{CO} and \textit{g}-\ce{O} + \textit{g}-\ce{NH2} $\rightarrow$ \textit{g}-\ce{H} + \textit{g}-\ce{HNO}.
    \item The abundance of \ce{NH2} increases due to the outburst through \textit{g}-\ce{CH3NH2+} + \ce{e-} $\rightarrow$ \textit{g}-\ce{NH2} + \textit{g}-\ce{CH3} and the same reactions that we retrieve for the 100 K increase: \textit{g}-\ce{NH4+} + \ce{e-} $\rightarrow$ \textit{g}-\ce{H2} (or \textit{g}-\ce{H} + \textit{g}-\ce{H}) + \textit{g}-\ce{NH2}.
    \item The \ce{HCNO} excess is due to the reaction \textit{g}-\ce{CH2} + \textit{g}-\ce{NO} $\rightarrow$ \textit{g}-\ce{HCNO} + \textit{g}-\ce{H}.
    \item The increase in the \ce{NO} abundance is produced through the reactions of H, C, and O atoms with \ce{NO2} and \ce{N2O}, with the addition of \textit{g}-\ce{N2O} $\rightarrow$ \textit{g}-\ce{N} + \textit{g}-\ce{NO}.
    \item Finally, the reaction \textit{g}-\ce{N} + \textit{g}-\ce{HCO} $\rightarrow$ \textit{g}-\ce{H} + \textit{g}-\ce{NCO} explains the higher gas phase abundance of \ce{NCO}.
\end{itemize}
The maximal amplitude of those bumps varies from one molecule to another. Compared to the abundance of the inner region, it goes from a factor of a few (\ce{NCO} and \ce{CH3O} for example) to more than one order of magnitude (\ce{HCCO}, \ce{HCNO}, and \ce{HNO}). Also, the time at which the maximum is reached is not the same for every molecule. For example, the excess of \ce{HCCO} reaches its maximum at the end of the outburst while the one of \ce{HCNO} is reached 5 kyr afterward (see Table \ref{tab:results_appendix}). This implies that the outburst does not affect the formation of the molecules in the same way. Depending on the species, it modifies the abundances of the key reactants at different locations within the envelope. The resulting chemical perturbation is then transported inward. The delay before the abundance excess reaches its maximum reflects the time required for the modified reactant abundances to propagate to the region where the molecule is efficiently formed. This explains why the timing of the maximum differs from one molecule to another. In addition, the outburst does not alter the abundances of the relevant reactants by the same amount for every species, leading to the observed variations in the amplitude of the abundance excesses.

\subsection{Long-term chemical impact}
\label{subsec:long_term_inner_impact}
Another impact we found in this study is that an outburst can modify the molecular abundances in the warm inner region of the envelope as late as 10--50 kyr after the outburst. These delayed changes are found for eight molecules: $\mathrm{HOCH_2CHO}$, $\mathrm{CH_3COCH_2OH}$, $\mathrm{C_2H_5CHO}$, $\mathrm{OCS}$, $\mathrm{SO_2}$, $\mathrm{C_2S}$, $\mathrm{C_3S}$, and $\mathrm{HC_3N}$ (see Table \ref{tab:results_appendix}).\\
Rows g and h of Fig. \ref{fig:radial_profile} show the radial abundance profiles of \ce{HOCH2CHO} and \ce{SO2}. Although they do not reveal a clear separation, there are certain differences between the inner quiescent abundance plateau and the outburst one ($T_\mathrm{quiescent}$>100 K) approximately 20 kyr after the outburst. These differences are more clearly shown in Fig. \ref{fig:inner_impact} and \ref{fig:inner_impact_ratio}, which are presented hereafter. For \ce{HOCH2CHO}, a decrease by a factor of 3 is observed in the gas phase abundance when comparing the outburst cases to the quiescent ones, whereas \ce{SO2} exhibits an enhancement by a factor of 2. These trends are also reflected in the ice-phase abundances (at larger radii) and in the total abundances.\\
To investigate in more detail the chemical evolution of the inner region between 10 and 50 kyr after the onset of the outburst, a set of supplementary models was computed for particles that terminate their trajectory at 10 au with $t_{post-trigger}$ times of 15, 25, 30, 35, 40, and 45 kyr. Figures \ref{fig:inner_impact} and \ref{fig:inner_impact_ratio} compile results from both the main model grid and the additional set of $t_{post-trigger}$, illustrating the chemical evolution of the inner region (at 10 au) for the eight molecules mentioned above, from the beginning of the outburst to 50 kyr afterward. A few tens of thousand years after the outburst, \ce{HOCH2CHO}, \ce{CH3COCH2OH}, \ce{C2H5CHO}, and \ce{OCS} are less abundant in the inner region than in the quiescent cases, whereas \ce{SO2}, \ce{C2S}, \ce{C3S}, and \ce{HC3N} are more abundant. For these eight molecules, the outburst alters the chemical composition of the cold outer envelope. As the impacted material continues its inward motion during the collapse, it transports this modified chemical imprint toward the inner regions, leading to a gas phase abundance discrepancy between outburst and quiescent cases.\\
For \ce{HOCH2CHO} and \ce{CH3COCH2OH} (\ce{C2H5CHO} and \ce{OCS}), the amplitude of the abundance discrepancy reaches a maximum 15--20 kyr (25--30 kyr) after the onset of the outburst and subsequently decreases with time. For \ce{SO2} (\ce{C2S} and \ce{C3S}), the abundance excess peaks at 20 kyr (25--30 kyr) after the outburst. For \ce{HC3N}, however, the abundance enhancement continues to grow at later time steps, suggesting that its maximum occurs beyond 50 kyr. The further away the chemical impact of the outburst occurs in the envelope, the later the difference in the inner abundance is seen.\\
In the case of \ce{HOCH2CHO}, its abundance decrease in the inner-region is brought by its diminished production on grains caused by the destruction of the icy reservoirs of colanic acid (\ce{HCO}) and hydroxymethyl (\ce{CH2OH}) in the outer envelope during the outburst, leading to a lower production of glycolaldehyde on grains. The chemical deficiency is carried inward as the particle migrates until it reaches the warm inner region, 20 kyr later, where its abundance is $\sim1$ order of magnitude lower than in the quiescent case.\\
The reduced availability of \ce{HCO} following the outburst also accounts for the lower inner abundances of \ce{CH3OCH2OH}, \ce{C2H5CHO}, and \ce{OCS}. For the other molecules:
\begin{itemize}
    \item \ce{SO2} is produced in greater amounts in the outburst case by destroying \textit{s}-\ce{HOSO}, and by transforming \textit{s}-\ce{SO} into \textit{s}-\ce{SO2} through reactions with \textit{s}-\ce{HOSO}, \textit{s}-\ce{NO2}, and \textit{s}-\ce{H2NO2}, whose abundances also increase during the outburst. Although its abundance is increasing, \ce{SO2} remains a secondary reservoir of sulfur.
    \item For \ce{C3S}, it is through the reactions \textit{g}-\ce{H} + \textit{g}-\ce{HC3S} $\rightarrow$ \textit{g}-\ce{H2} + \textit{g}-\ce{C3S} and \textit{s}-\ce{S} + \textit{s}-\ce{C3} $\rightarrow$ \textit{s}-\ce{C3S} (or \textit{g}-\ce{C3S}). Indeed, the outburst enhances the production of \ce{HC3S} and facilitates the reactions consuming \ce{C3}.
    \item The \ce{C2S} higher abundance is linked to the \ce{C3S} and \ce{HC3S} increases, with the reactions \textit{g}-\ce{O} + \textit{g}-\ce{C3S} $\rightarrow$ \textit{g}-\ce{CO} + \textit{g}-\ce{C2S}, \textit{g}-\ce{O} + \textit{g}-\ce{HC3S} $\rightarrow$ \textit{g}-\ce{HCO} + \textit{g}-\ce{C2S}, and \textit{g}-\ce{N} + \textit{g}-\ce{HC3S} $\rightarrow$ \textit{g}-\ce{HCN} + \textit{g}-\ce{C2S}.
    \item The \ce{HC3N} excess is produced via several reactions consuming \ce{HC3NH+}, \ce{C2H3CNH+}, \ce{H2C3N}, and \ce{C3N} after the outburst.
\end{itemize}
\begin{figure*}
    \centering
    \includegraphics[width=0.80\textwidth]{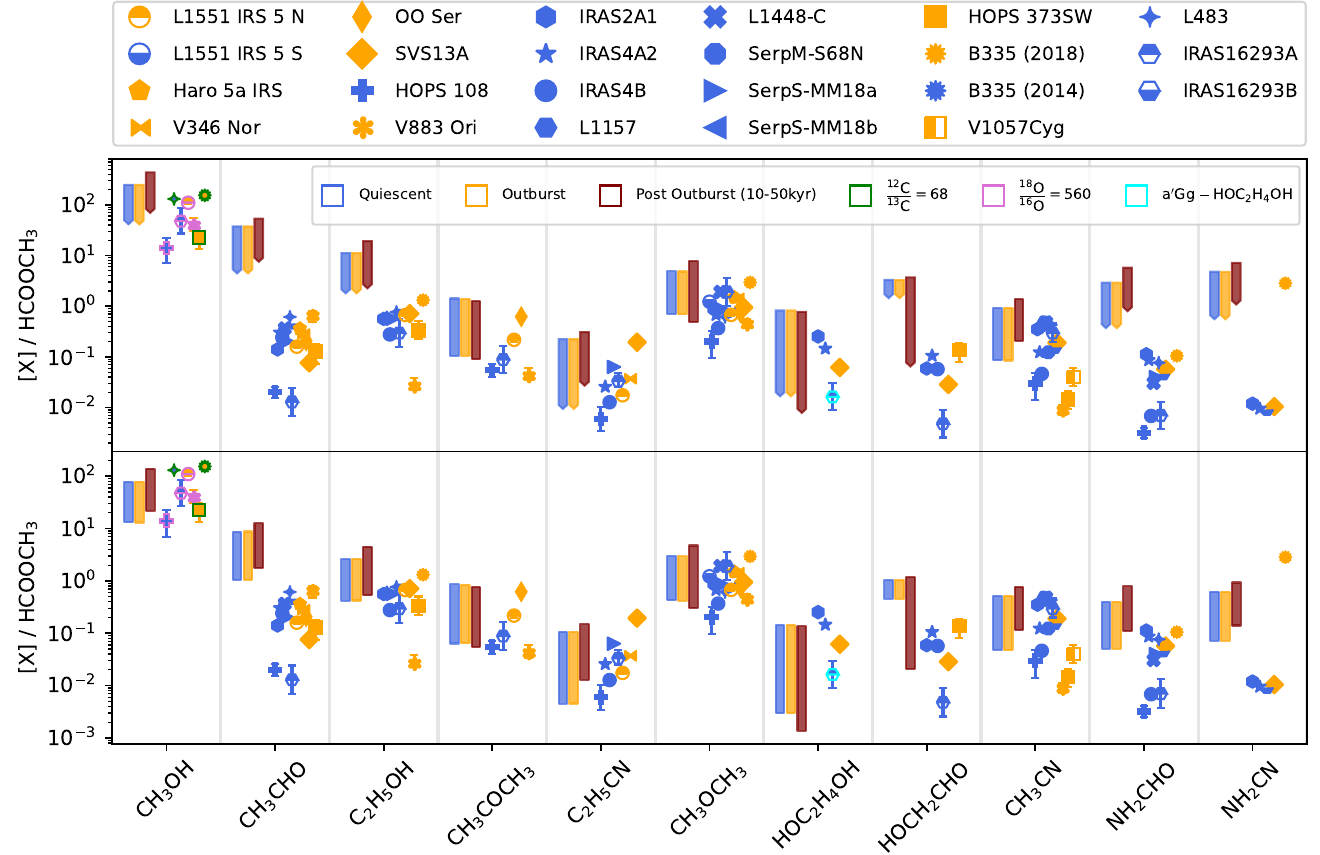}
    \caption{Comparison between predictions of our model and observational data of low-mass protostars from the literature. Model predictions (vertical bars) are obtained by taking the abundances at 10 au as representative of the abundance plateau of molecules, while observational data from the literature are represented by points. The methanol points indicated with green and purple edges indicate that they were obtained using its $\mathrm{^{13}C}$ and $\mathrm{^{18}O}$ isotopologs. For \ce{HOC2H4OH}, we included a data point, with a cyan border, that only includes the conformer a'Gg-\ce{HOC2H4OH}. Top: Raw model ratios. Arrows are used for the abundance ratios of molecules that present different spatial extents. Indeed, the observations assume similar source sizes to derive the abundances of the molecules. Bottom: Ratios corrected for differences in emission region sizes (see explanations in Sect. \ref{subsubsec:obs_comp}).}
    \label{fig:comparison_obs}
\end{figure*}

\section{Discussion} \label{sec:discussion}

\subsection{Comparison with previous studies}

\begin{figure}
    \centering
    \includegraphics[width=7cm]{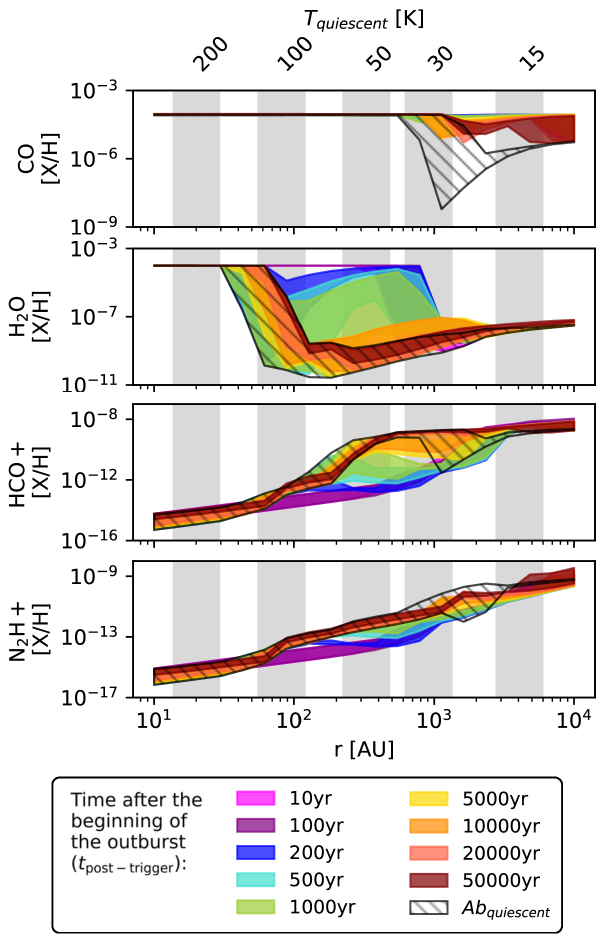}
    \caption{Radial gas phase abundance profiles of \ce{CO}, \ce{H2O}, \ce{HCO+}, and \ce{N2H+}.}
    \label{fig:N2H+_HCO+}
\end{figure}

\subsubsection{COMs in the warm inner regions of protostars}
\label{subsubsec:obs_comp}
To assess whether our model is consistent with observations of COMs toward low-mass protostars, we compiled column densities from interferometric studies found in the literature. As interferometric studies probes the most compact regions, i.e., the warm inner region of the envelope (hot corino), we compared the retrieved column densities with our model predictions of this region. We used the same list of sources and papers as those presented in \cite{Cruz2025}, with the addition of IRAS4A2, IRAS4B, L1157, SerpS-MM18b, and V1057 Cyg. In summary, this includes L1551 IRS 5 N, L1551 IRS 5 S, Haro 5a IRS, V346 Nor, and OO Ser \citep{Cruz2025}, SVS13A, IRAS2A1, IRAS4A2, IRAS4B, L1157, L1448-C, SerpM-S68N, SerpS-MM18a, and SerpS-MM18b \citep{Belloche2020}, HOPS 108 \citep{Chahine2022}, V883 Ori \citep{Jeong2025}, HOPS 373SW \citep{Lee2023}, B335 \citep{Lee2025}, V1057 Cyg \citep{Calahan2024}, L483 \citep{Jacobsen2019}, IRAS16293A \citep{Manigand2020, Calcutt2018}, and IRAS16293B \citep{Jorgensen2018, Coutens2016, Coutens2018, Calcutt2018}. These objects span a wide range of luminosities, evolutionary stages, and envelope masses, providing a representative view of the chemical diversity observed in low-mass protostars. Figure \ref{fig:comparison_obs} presents a comparison between our model predictions (at 10 au) and observational data from the previous studies. Given that the presented species exhibit an abundance plateau in the inner region, we consider the abundance at 10 au to be representative of the entire desorption region. To compute the abundance ratios from our models presented in Fig. \ref{fig:comparison_obs}, the abundances of the two involved molecules were extracted from the last time step of the same model, which is the moment when the particle in the selected models reaches 10 au. In this figure, \ce{HCOOCH3} (methyl formate) is used as the reference molecule. Fig. \ref{fig:comp_obs_first} shows similar plots with other molecules as reference. We indicated sources that are thought to be in a quiescent phase in blue, and those undergoing an outburst in orange. The same color scheme was applied to the models (vertical bars), with the addition of red, corresponding to the predicted abundances in the outburst case from 10 to 50 kyr after the outburst occurred. In the upper panel, we display the abundance ratios computed from our grid of model (all $t_\mathrm{trigger}$ combined) as well as the column density ratios derived from observational data. In the lower panel, we correct our abundance ratios to account for differences in the sizes of the emitting regions, i.e., the regions of a high gas phase abundance. Indeed, molecules are emitted in regions of different sizes, and the radiative transfer models, used to retrieve column densities, assume that they all come from regions of the same size. Therefore, for a relevant comparison, a correction needs to be applied to the modeled abundance ratios. In the upper panel, arrows mark the direction of the correction applied to the model predictions (vertical bars), for cases in which the correction changes the ratios by more than a factor of two. The correction consists of using
\begin{equation}
    \mathrm{\frac{N_X}{N_Y}=\frac{[X]}{[Y]}\ \frac{(r_X^{\frac{3}{2}}-1)}{(r_Y^{\frac{3}{2}}-1)}}
    \label{eq:obs_corr}
\end{equation}
instead of $\frac{[\mathrm{X}]}{[\mathrm{Y}]}$. [X] and [Y] are the abundances of species X and Y predicted by our model at 10 au, and $\mathrm{r_{X}}$ and $\mathrm{r_{Y}}$ are the radii of the emission size (in astronomical units) predicted by our model for species X and Y. Appendix \ref{app:corr} presents how we obtain this correction, while Appendix \ref{app:rX_rY} explains how we determined $\mathrm{r_{X}}$ and $\mathrm{r_{Y}}$.\\
Overall, the corrected abundance ratios reproduce the observed trends within one order of magnitude, consistent with the intrinsic dispersion among sources. We do not see any significant differences between the abundance ratios in the quiescent case and during the outburst, both for models and observations (see also \citealp{Cruz2025}). In the inner regions, luminosity outbursts do not substantially alter the abundance ratios of COMs in the scale of the outburst's lifetime. The dispersion observed among sources could be related to differences in protostellar age, as suggested by the spread indicated by the vertical bars representing our model predictions. Other factors, such as the characteristics of the observed system or its surrounding environment, may also contribute to the observed dispersion.\\
Some discrepancies between the model predictions and the observations are observed in Figures \ref{fig:comparison_obs} and \ref{fig:comp_obs_first}, even after the correction for the different emission sizes. This is especially the case for \ce{CH3CHO}, \ce{NH2CHO}, and \ce{NH2CN} when compared to the other COMs. This could be related to the binding energies assumed in our network. Indeed, these three species have the highest binding energies, with $\mathrm{E_{bind,\ce{CH3CHO}}}=5500$ K, $\mathrm{E_{bind,\ce{NH2CHO}}}=6500$ K, and $\mathrm{E_{bind,\ce{NH2CN}}}=6600$ K, while the values are lower for the other COMs (e.g., $\mathrm{E_{bind},\ce{HCOOCH3}}=3500$ K). These binding energies were computed using the methodology presented in \citet{Wakelam2017} and are consistent with values reported in the literature \citep[e.g.,][]{Ferrero2022,Minissale2022}. If the binding energies of \ce{CH3CHO}, \ce{NH2CHO}, and \ce{NH2CN} were underestimated, the correction would not be sufficient. However, the reason for this discrepancy could also be that in the innermost regions, the spherical assumption used for the correction is not valid.\\
The observed \ce{HOCH2CHO} abundances are lower than those predicted by our model in the quiescent case and during the outburst but better match the abundance predicted a few tens of thousand years after the outburst. This might indicate a delayed chemical response, possibly resulting from a past outburst that occurred a couple tens of thousand years ago (see Sect. \ref{subsec:long_term_inner_impact}). In this case, the observed protostars of this sample would have experienced a luminosity outburst approximately 15--25 kyr ago, considering a similar outburst and protostellar luminosity. Another possibility that we cannot exclude is that reactions in the chemical network are missing or erroneous for this species.\\
The abundance ratios of other COMs are different 10 to 50 kyrs after the outburst (vertical red bars in Fig. \ref{fig:comparison_obs}) compared to both the quiescent case and the outburst phase (blue and orange vertical bars, respectively). These species are not classified as impacted on long-term timescales in Sect. \ref{subsec:long_term_inner_impact}, as their post-outburst abundances (relative to H) remain within the range of quiescent values, even though their ratios with respect to other COMs can vary. Nevertheless, these variations are clearly smaller than that observed for glycolaldehyde.

\subsubsection{Expansion of snowlines and associated chemical signatures}
As discussed in Sect. \ref{sec:results}, the luminosity outburst pushes the molecular snowlines further into the envelope. For most species, the radial extent of the abundance plateau during the outburst becomes approximately one order of magnitude larger than in the quiescent phase. In particular, \ce{H2O} and CO exhibit spatially extended gas-phase regions during the outburst, consistently with previous studies \citep{Visser2015, Jorgensen2015, Hsieh2019, Frimann2017, Narayanan2023}. For \ce{H2O}, we recover both a snowline expansion factor of $\sim10$ and a refreeze timescale of $\sim10^3$ yr after the outburst, in agreement with the results of \citet{Visser2015}.\\
The relocation of the \ce{H2O} snowline is expected to affect the chemistry of \ce{HCO+} and \ce{N2H+}. Previous studies proposed that gaseous \ce{H2O} efficiently destroys these ions through gas-phase reactions, making \ce{HCO+} and \ce{N2H+} potential tracers of the \ce{H2O} sublimation region \citep{Jorgensen2013, Visser2015, Hsieh2019, vanthoff2018}. In our model, we also find that the abundances of \ce{HCO+} and \ce{N2H+} decrease in the region where \ce{H2O} sublimates during the outburst (Fig. \ref{fig:N2H+_HCO+}). However, the dominant destruction pathways differ from the commonly assumed picture. For \ce{HCO+}, reactions with gaseous \ce{H2O} contribute to the depletion, but the dominant destruction pathway in our chemical network is
\begin{equation}
\ce{H2CO + HCO+ -> CO + H2COH+}.
\end{equation}
For \ce{N2H+} where \ce{H2O} sublimates, the dominant destruction pathway is instead
\begin{equation}
\ce{CO + N2H+ -> N2 + HCO+},
\end{equation}
which becomes more efficient during the outburst.\\
Previous studies also suggested that the outward expansion of the CO snowline should produce an enhancement of \ce{HCO+} and a depletion of \ce{N2H+} \citep{Visser2015, Hsieh2019, Sharma2020}. Our results qualitatively agree with these trends. However, the corresponding variations remain weak in our abundance profiles and do not produce a significant chemical transition at the CO sublimation front. This result is consistent with previous works showing that the relationship between \ce{N2H+} and the CO snowline can be more complex than often assumed \citep{vanthoff2017, Murillo2022}.

\subsection{Model limitations} \label{subsec:model_limitations}
In this study, we adopted fixed values for the mass of the infalling cloud, for its initial temperature, and for the visual extinction of the surrounding molecular cloud. However, the chemical outcomes may differ for systems of different masses, for those exposed to strong irradiation from nearby massive stars, or for less embedded systems (see Appendix \ref{subsec:UV}). Furthermore, our model does not include UV photons emitted by the protostar and its accretion region, which is a reasonable assumption for radii $\geq$10 au, as in our case (see Appendix \ref{subsec:UV}).\\
We assume a thermal equilibrium between gas and dust ($T_\mathrm{gas} = T_\mathrm{dust}$) throughout the envelope. This approximation is valid in regions with densities above $\approx10^5~\mathrm{cm^{-3}}$, where collisional coupling efficiently equalizes their temperatures \citep{Goldsmith2001}. In the outermost envelope, where densities drop to $\sim 10^4~\mathrm{cm^{-3}}$, the gas may be a few kelvins warmer than the dust. We assume that this temperature difference has a negligible impact on the overall chemical evolution in our model. In addition, we assume spherical dust grains with a constant grain size, a simplification that can influence ice chemistry by altering the available surface area for grain-surface reactions \citep{Ge2016, Pauly2016}.\\
Finally, our model is one-dimensional, and focuses on the protostellar envelope only. This approach neglects structures such as accretion streams, disks, or outflows, whose feedback can locally alter the physical conditions and, consequently, the chemistry.

\section{Conclusion} \label{sec:conclusion}

We have modeled the chemical response of a collapsing protostellar envelope to a luminosity outburst representative of a FUor accretion event. Using a dynamic physico–chemical framework combining the APE and NAUTILUS codes, we investigated the evolution of abundances for 83 molecules across a broad range of spatial and temporal scales.\\
Our results show that luminosity outbursts strongly alter the molecular distribution within the envelope, primarily by shifting snowlines outward by up to an order of magnitude in radius. This leads to an enrichment of the gas phase by COMs and other volatile species. The time needed for species to return to their pre-outburst abundances varies from a few hundred years in the inner 100 au to several tens of thousand years beyond 1000 au. We identified eight molecules whose inner abundances are affected tens of thousand years after the event: \ce{HOCH2CHO}, \ce{CH3COCH2OH}, \ce{C2H5CHO}, OCS, \ce{SO2}, \ce{C2S}, \ce{C3S}, and \ce{HC3N}. This long-term impact arises from chemical changes induced by the outburst in the cold outer envelope, which is subsequently carried inward during the collapse. We also found that \ce{CH3O}, \ce{C2O}, \ce{NH2}, \ce{HCCO}, \ce{HCNO}, \ce{HNO}, \ce{NO}, and \ce{NCO} show a bump in their gas-phase abundance profile between $\sim$ 300 au and $\sim$ 1500 au in response to the outburst.\\
The model reproduces several observational trends reported in young protostars, such as the outward motion of CO and \ce{H2O} snowlines as well as the associated abundance profiles of \ce{N2H+} and \ce{HCO+}. It also predicts abundance ratios between COMs that are relatively consistent with interferometric observations of low-mass protostars.

\begin{acknowledgements}
This study is part of a project that has received funding from the European Research Council (ERC) under the European Union’s Horizon 2020 research and innovation programme (Grant agreement No. 949278, Chemtrip). Valentine Wakelam thanks the CNRS program "Physique et Chimie du Milieu Interstellaire" (PCMI) co-funded by the Centre National d’Etudes Spatiales (CNES).
\end{acknowledgements}

\bibliography{bib}
\bibliographystyle{aa}

\appendix
\onecolumn
\section{Table summarizing the impacts of outbursts for each molecule}

\begin{table*}[!ht]
\caption{Summary of impacts related to the luminosity outburst for the selected set of molecules.}
\label{tab:results_appendix}
\centering
\begin{tabular}{r|cccccc|c|c|c}
\hline
\hline
 & \multicolumn{6}{c|}{\textbf{Abundance plateau\tablefootmark{a}}} & \textbf{Abundance} & \textbf{Long-term}\\
 & \multicolumn{6}{c|}{} & \textbf{bump (300-1500 au)\tablefootmark{b}} & \textbf{inner impact\tablefootmark{c}}\\
\textbf{Molecule}  & \textbf{TD/CR\tablefootmark{0}} & \textbf{Ab plat\tablefootmark{1}} & \textbf{r$_\mathrm{q}$\tablefootmark{2}} & \textbf{r$_\mathrm{o}$\tablefootmark{3}} & \textbf{T$_\mathrm{qq}$\tablefootmark{4}} & \textbf{T$_\mathrm{qo}$\tablefootmark{5}} & \textbf{t$_\mathrm{bump}$} & \textbf{t$_\mathrm{long\ term\ inner}$} & \textbf{Ab max\tablefootmark{d}}\\
 & & (/H) & (au) & (au) & (K) & (K) & (yr) & (kyr) & (/H)\\
\hline
\ce{C2H3CHO} & TD &  1.0E-08 & 79 & 798 & 111 & 35    & - & -    & 4.6E-08 \\
\ce{C2H5CN}  & TD  & 6.6E-09 & 128 & 1287 & 88 & 27   & - & -    & 1.4E-08 \\
\ce{C2H5OCH3} & TD & 1.3E-08 & 97 & 977 & 101 & 31    & - & -    & 2.0E-08 \\
\ce{C2H5OH}   & TD & 5.1E-07 & 79 & 798 & 111 & 35    & - & -    & 6.2E-07 \\
\ce{C3H6}    & TD  & 1.4E-07 & 197 & 1975 & 71 & 22   & - & -    & 2.1E-07 \\
\ce{C3O}      & TD & 5.8E-08 & 157 & 1576 & 79 & 25   & - & -    & 2.4E-07 \\
\ce{CH2NH}    & TD & 4.9E-08 & 97 & 977 & 101 & 31    & - & -    & 7.5E-08 \\
\ce{CH3CCH}   & TD & 3.1E-07 & 157 & 1576 & 79 & 25   & - & -    & 4.2E-07 \\
\ce{CH3CHO}   & TD & 1.6E-06 & 79 & 798 & 111 & 35    & - & -    & 1.7E-06 \\
\ce{CH3CN}    & TD & 3.2E-08 & 141 & 1416 & 83 & 26   & - & -    & 4.9E-08 \\
\ce{CH3COCH3} & TD & 4.1E-08 & 150 & 1506 & 81 & 25   & - & -    & 6.3E-08 \\
\ce{CH3COOH}  & TD & 2.9E-08 & 65 & 658 & 123 & 38    & - & -    & 4.7E-08 \\
\ce{CH3Cl}    & TD & 8.9E-10 & 175 & 1755 & 75 & 23   & - & -    & 1.0E-09 \\
\ce{CH3NC}    & TD & 1.3E-10 & 157 & 1576 & 79 & 25   & - & -    & 1.6E-10 \\
\ce{CH3NCO}   & TD & 6.5E-09 & 111 & 1117 & 94 & 29   & - & -    & 2.9E-08 \\
\ce{CH3OCH2OH} & TD & 4.4E-09 & 65 & 658 & 123 & 38    & - & -    & 3.9E-08 \\
\ce{CH3OCH3}  & TD & 2.1E-07 & 150 & 1506 & 81 & 25   & - & -    & 2.6E-07 \\
\ce{CH3OH}    & TD & 1.2E-05 & 97 & 977 & 101 & 31    & - & -    & 1.4E-05 \\
\ce{CH3SH}    & TD & 1.4E-08 & 149 & 1496 & 81 & 25   & - & -    & 3.5E-08 \\
\ce{CH4}      & TD & 1.5E-05 & 3271 & 32663 & 17 & 5 & - & -    & 1.6E-05 \\
\ce{CNCN}     & TD & 2.9E-12 & 157 & 1576 & 79 & 25   & - & -    & 6.7E-12 \\
\ce{CO}       & TD & 8.5E-05 & 1721 & 17189 & 24 & 7 & - & -    & 9.2E-05 \\
\ce{CO2}      & TD & 4.5E-06 & 468 & 4679 & 46 & 14   & - & -    & 5.0E-06 \\
\ce{H2CCO}    & TD & 2.4E-06 & 394 & 3940 & 50 & 15   & - & -    & 2.9E-06 \\
\ce{H2CO}     & TD & 2.4E-05 & 253 & 2534 & 62 & 19   & - & -    & 3.1E-05 \\
\ce{H2CS}     & TD & 2.2E-08 & 141 & 1416 & 83 & 26   & - & -    & 3.0E-08 \\
\ce{H2O}      & TD & 9.9E-05 & 76 & 768 & 114 & 36    & - & -    & 1.0E-04 \\
\ce{HCOOCH3}  & TD & 8.4E-08 & 209 & 2095 & 69 & 21   & - & -    & 5.7E-07 \\
\ce{HOC2H4OH} & TD & 3.1E-08 & 65 & 658 & 123 & 38    & - & -    & 3.0E-07 \\
\ce{HOCH2CN}  & TD & 8.1E-10 & 157 & 1576 & 79 & 25   & - & -    & 1.2E-09 \\
\ce{HOCHCHCHO} & TD & 2.6E-08 & 65 & 658 & 123 & 38    & - & -    & 1.0E-07 \\
\ce{HONO}     & TD & 2.1E-08 & 41 & 419 & 154 & 48    & - & -    & 3.1E-08 \\
\ce{N2O}      & TD & 1.9E-07 & 337 & 3372 & 54 & 17   & - & -    & 2.0E-07 \\
\ce{NH2CHO}   & TD & 1.3E-07 & 55 & 558 & 133 & 42    & - & -    & 1.6E-07 \\
\ce{NH2CN}    & TD & 1.8E-07 & 53 & 538 & 136 & 43    & - & -    & 2.1E-07 \\
\ce{NH3}      & TD & 3.3E-06 & 97 & 977 & 101 & 31    & - & -    & 3.3E-06 \\
\ce{NS}       & TD & 2.7E-10 & 312 & 3122 & 56 & 17    & - & -    & 1.2E-09 \\
\ce{c-C2H4O}  & TD & 2.2E-08 & 97 & 977 & 101 & 31    & - & -    & 2.7E-08 \\
\ce{l-C3H2}   & TD & 3.0E-10 & 116 & 1167 & 92 & 29   & - & -    & 9.5E-10 \\
\ce{c-C3H2}   & TD & 8.8E-09 & 68 & 688 & 120 & 38   & - & -  & 1.3E-07 \\
\ce{HCOOH}    & CR & 5.9E-07 & 42 & 379 & 153 & 51    & - & -    & 8.5E-07 \\
\ce{CH2OH}   & CR  & 1.8E-08 & 379 & 3359 & 51 & 17   & - & -    & 4.7E-08 \\
\ce{H2S}      & CR & 3.7E-08 & 61 & 784 & 128 & 35    & - & -    & 1.1E-07 \\
\ce{HNC}      & CR & 3.2E-09 & 61 & 545 & 128 & 43    & - & -    & 3.8E-08 \\
\ce{HCN}      & CR & 3.2E-07 & 61 & 784 & 128 & 35    & - & -    & 4.4E-07 \\
\ce{HNCO}     & CR & 1.6E-07 & 42 & 379 & 153 & 51    & - & -    & 2.2E-07 \\
\ce{C2H5CHO}  & TD & 2.0E-08 & 122 & 1227 & 90 & 28   & - & 40    & 2.3E-08 \\
\ce{OCS}      & TD & 9.7E-08 & 468 & 4679 & 46 & 14   & - & 25    & 1.3E-07 \\
\ce{HOCH2CHO}  & TD & 1.9E-07 & 97 & 977 & 101 & 31  & - & 20     & 1.3E-06 \\
\ce{SO2}       & TD & 8.4E-09 & 222 & 2224 & 66 & 21 & - & 20     & 3.4E-08 \\
\ce{CH3COCH2OH} & TD & 3.0E-08 & 65 & 658 & 123 & 38  & - & 20     & 1.2E-07 \\
\ce{HC3N}      & TD & 1.4E-09 & 157 & 1576 & 79 & 25 & - & 50     & 1.8E-08 \\
\ce{C3S}       & TD & 2.0E-09      & 157 & 1576 & 79 & 25 & - & 30     & 9.2E-09 \\
\hline
\end{tabular}
\end{table*}

\begin{table*}[]
\centering
\caption{Continuation of Table \ref{tab:results_appendix}}
\begin{tabular}{r|cccccc|c|c|c}
\hline
\hline
 & \multicolumn{6}{c|}{\textbf{Abundance plateau\tablefootmark{a}}} & \textbf{Abundance} & \textbf{Long-term}\\
 & \multicolumn{6}{c|}{} & \textbf{bump (300-1500 au)\tablefootmark{b}} & \textbf{inner impact\tablefootmark{c}}\\
\textbf{Molecule}  & \textbf{TD/CR\tablefootmark{0}} & \textbf{Ab plat\tablefootmark{1}} & \textbf{r$_\mathrm{q}$\tablefootmark{2}} & \textbf{r$_\mathrm{o}$\tablefootmark{3}} & \textbf{T$_\mathrm{qq}$\tablefootmark{4}} & \textbf{T$_\mathrm{qo}$\tablefootmark{5}} & \textbf{t$_\mathrm{bump}$} & \textbf{t$_\mathrm{long\ term\ inner}$} & \textbf{Ab max\tablefootmark{d}}\\
 & & (/H) & (au) & (au) & (K) & (K) & (yr) & (kyr) & (/H)\\
\hline
\ce{C2S}   & -    & -       & - & - & -           & - & - & 25     & 3.3E-10 \\
\ce{C2O}    & -  & - & - & - & - & -   & 10-5000 (1000) & - & 5.4E-09 \\
\ce{NH2}     & - & - & - & - & - & -    & 10-1000 (1000) & - & 1.6E-08 \\
\ce{NCO}  & -    & - & - & - & - & -   & 10-1000 (1000) & -    & 2.2E-09 \\
\ce{NO}    & -    & -       & - & - & -               & - & 10-5000 (100)   & -     & 5.3E-08 \\
\ce{HNO}    & -   & -       & - & - & -               & - & 10-5000 (500-1000)  & -   & 4.6E-09 \\
\ce{HCCO}    & -  & -       & - & - & -               & - & 10-1000 (100)  & -  & 2.6E-08 \\
\ce{HCNO}     & - & -       & - & - & -               & - & 10-10000 (5000)  & - & 7.9E-10 \\
\ce{CH3O}    & - & - & - & - & - & -    & 10-5000 (1000) & - & 1.3E-09 \\
\ce{C2H}      & - & -       & - & - & -               & - & -   & -     & 5.0E-08 \\
\ce{C3N}      & - & -       & - & - & -               & - & -   & -     & 2.0E-10 \\
\ce{C4H}      & - & -       & - & - & -               & - & -   & -     & 5.4E-09 \\
\ce{CH2CN}    & - & -       & - & - & -               & - & -   & -     & 1.9E-08 \\
\ce{CH3+}     & - & -       & - & - & -               & - & -   & -     & 4.6E-10 \\
\ce{CN}        & - & -       & - & - & -               & - & -   & -     & 2.4E-08 \\
\ce{CS}        & - & -       & - & - & -               & - & -   & -     & 6.3E-09 \\
\ce{H2CN}     & - & -       & - & - & -               & - & -   & -     & 8.2E-10 \\
\ce{H2NC}     & - & -       & - & - & -               & - & -   & -     & 5.1E-11 \\
\ce{H3+}      & - & -       & - & - & -               & - & -   & -     & 1.1E-08 \\
\ce{H3O+}     & - & -       & - & - & -               & - & -   & -     & 1.1E-09 \\
\ce{HCO}       & - & -       & - & - & -               & - & -   & -     & 2.4E-07 \\
\ce{HCS}       & - & -       & - & - & -               & - & -   & -     & 3.5E-09 \\
\ce{HOCN}      & - & -       & - & - & -               & - & -   & -     & 7.9E-12 \\
\ce{HOCO+}     & - & -       & - & - & -               & - & -   & -     & 5.8E-11 \\
\ce{HSC}       & - & -       & - & - & -               & - & -   & -     & 4.2E-12 \\
\ce{N2H+}     & - & -       & - & - & -               & - & -   & -     & 3.5E-09 \\
\ce{NH}       & - & - & - & - & - & -   & - & -    & 2.0E-08 \\
\ce{OH}       & - & -       & - & - & -               & - & -   & -     & 1.5E-07 \\
\ce{SO}       & - & -       & - & - & -               & - & -   & -     & 7.8E-10 \\
\ce{c-C3H}   & -  & -       & - & - & -               & - & -   & -     & 7.6E-10 \\
\hline
\end{tabular}
\tablefoot{
\tablefoottext{a}{Characteristics of the abundance plateau (see Sect. \ref{subsec:inner_spatial_variations} and \ref{subsubsec:other_processes}). If the molecule does not present a plateau, the columns are filled with "-". \tablefoottext{0}{Nature of the spatial variation of the abundance plateau. TD corresponds to direct thermal desorption (see Sect. \ref{subsec:inner_spatial_variations}). CR indicates that the plateau is produced by chemical reactions (see Sect. \ref{subsubsec:other_processes}).} \tablefoottext{1}{Mean gas phase abundance of the plateau taken at 10 au, relative to H.} \tablefoottext{2}{r$_\mathrm{q}$ denotes the outer radii of the abundance plateau in the quiescent state.} \tablefoottext{3}{r$_\mathrm{o}$ denotes the outer radii of the abundance plateau in the outburst state.} \tablefoottext{4}{T$_\mathrm{qq}$ is the corresponding quiescent temperatures at r$_\mathrm{q}$.} \tablefoottext{5}{T$_\mathrm{qo}$ is the corresponding quiescent temperatures at r$_\mathrm{o}$.}}
\tablefoottext{b}{Range of time ($t_\mathrm{post-trigger}$) during which the abundance bump seen between $\sim300$ and $\sim1500$ au exists. The time at which it is the most pronounced is indicated in parentheses (see Sect. \ref{subsec:outer_spatial_variations}). If the molecule does not present this bump, the columns are filled with "-".}
\tablefoottext{c}{Time ($t_\mathrm{post-trigger}$) at which the long-term inner impact is the most pronounced (see Sect. \ref{subsec:long_term_inner_impact}). If the molecule does not present this impact, the columns are filled with "-".}
\tablefoottext{d}{Maximal abundance of the molecule (relative to H), all models combined (radius and time).}
}
\end{table*}

\twocolumn
\clearpage
\section{Sketch of the physicochemical model}

\begin{figure}[H]
    \begin{minipage}[c]{\textwidth}
    \centering
    \includegraphics[width=14.5cm]{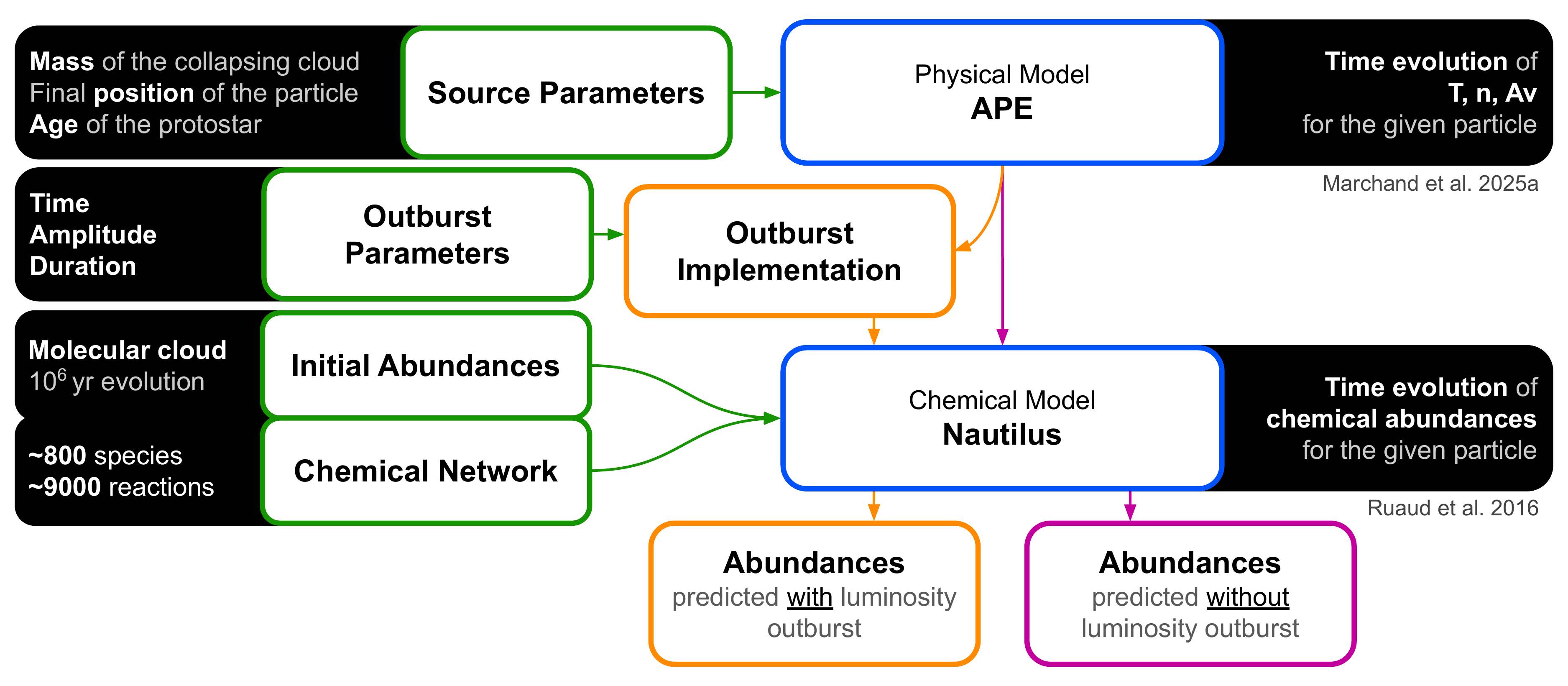}
    \caption{Sketch of the physicochemical model used in this study. The blue blocks represent the physical and the chemical models. The orange color refers to the outburst models, while the magenta color is for the quiescent models. The green blocks refer to the inputs of the model.}
    \label{fig:diagram}
    \end{minipage}
\end{figure}

\section{Comparison with observations: Correction of abundance ratios}
\label{app:corr}
To compare the abundance ratios predicted by our model with the observed column density ratios, we must account for the difference in emission size between the two species. The total number of observed molecules for the species X is
\begin{equation}
 \mathrm{Nb_{obs}(X) = N_X\ \pi\ R_X^2}
\end{equation}
where $\mathrm{N_X}$ is the column density of the molecule X and $\mathrm{R_X}$ the radius of its observed emission region.\\
From the models, the total number of molecules present in the emitting region is
\begin{equation}
 \mathrm{Nb_{model(X)} = \int_{r_0}^{r_X} [X]\ 4\ \pi\ r^2\ n_H(r)\ dr}
\end{equation}
with
\begin{equation}
    \mathrm{n_H(r) = n_H(r_{10au})\ (\frac{r}{r_{10au}})^{-\frac{3}{2}}}
\end{equation}
where [X] is the abundance of the molecule X relative to hydrogen, $\mathrm{r_X}$ the radius of the emitting region predicted by the model and $\mathrm{n_H}$ the density of H. We suppose a 1 au gap between the surface of the protostar and the inner part of the envelope, therefore $r_0$ is set to 1 au. We have :
\begin{equation}
    \mathrm{Nb_{model(X)} = [X]\ \frac{8}{3}\ \pi\ n_H(r_{10au})\ r_{10au}^{\frac{3}{2}}\ (r_X^{\frac{3}{2}}-r_0^{\frac{3}{2}})}
\end{equation}
Since $\mathrm{\frac{Nb_{obs}\ (X)}{Nb_{obs}\ (Y)}}$=$\mathrm{\frac{Nb_{model}\ (X)}{Nb_{model}\ (Y)}}$, we have:
\begin{equation}
    \mathrm{\frac{N_X}{N_Y}=\frac{[X]}{[Y]}\ \frac{(r_X^{\frac{3}{2}}-r_0^{\frac{3}{2}})}{(r_Y^{\frac{3}{2}}-r_0^{\frac{3}{2}})}\ \left(\frac{R_Y}{R_X}\right)^2}
\end{equation}\\
In the observations we compiled, X and Y in every column density ratios have the same estimated emission size.\\
We therefore have $\mathrm{\frac{R_Y}{R_X}} = 1$. We end up with:
\begin{equation}
    \mathrm{\frac{N_X}{N_Y}=\frac{[X]}{[Y]}\ \frac{(r_X^{\frac{3}{2}}-r_0^{\frac{3}{2}})}{(r_Y^{\frac{3}{2}}-r_0^{\frac{3}{2}})}}
\end{equation}\\

\vspace{23em}
\section{Derivation of the radius of the emission size, $\mathrm{r_{X}}$}
\label{app:rX_rY}
To derive the radius of emission size of a molecule $i$, we computed the radius at which the molecule thermally desorbs. To do so, we compute the rate coefficients for thermal desorption, $k_\mathrm{des}$, and for accretion, $k_\mathrm{acc}$. We use the description provided in \citet{Ruaud2019}:
\begin{equation}
    k_\mathrm{acc} (T) = \alpha_i v_i \langle \sigma_d n_d \rangle
\end{equation}
and
\begin{equation}
    k_\mathrm{des}(i) (T)= \nu_i\ exp \left( - \frac{E_\mathrm{bind}(i)}{T}\right)
\end{equation}
where
\[
\nu_i = \sqrt{\frac{2 N_\mathrm{sites} k E_\mathrm{bind}(i)}{\pi^2 m_i}} .
\]
$\alpha_i$ is the sticking probability of the species (that is assumed to be equal to 1 for all the species), $v_i$ its thermal speed, $\langle \sigma_d n_d \rangle$ the product of the dust cross-sectional area and the dust density. $\nu_i$ is the vibrational frequency of the species $i$, $E_\mathrm{bind}$ is the binding energy, $m_i$ is the mass, and $N_\mathrm{sites} \sim 10^{15}\ \mathrm{cm^{-2}}$ is the number of adsorption sites per $\mathrm{cm^2}$.\\
\\
For all the species, at low temperature, $k_\mathrm{acc}$$\gg$$k_\mathrm{des}$, and at high temperature, $k_\mathrm{acc}$$\ll$$k_\mathrm{des}$. The desorption temperature, $T_\mathrm{des}$, corresponds to the temperature at which $k_\mathrm{des}$ starts to overcome $k_\mathrm{acc}$. We thus take the temperature that satisfies $k_\mathrm{des} = k_\mathrm{acc}$ as the desorption temperature. Finally, for a given species $X$, the radius of its emission size, $r_{X}$, is the radius at which its desorption temperature is reached in our envelope model:
\begin{equation}
    r_{X} = \frac{T_\mathrm{ps}^{2}R_\mathrm{ps}}{\sqrt{T_\mathrm{des}^{4}-T_\mathrm{mc}^{4}}} .
\end{equation}
The values computed for the COMs studied here are consistent with the snowline positions seen in Fig. \ref{fig:radial_profile} for \ce{CH3OH} and on \url{https://zenodo.org/records/23667779} for the other COMs.

\section{UV irradiation} \label{subsec:UV}
In our model, the protostellar system is embedded in a molecular cloud that shields the interstellar radiation field, corresponding to an extinction of $A_\mathrm{V\ mc} = 5~\mathrm{mag}$ (see Table~\ref{tab:parameters}). It is therefore important to note that the results for the outer, cold envelope would differ for systems that are less embedded in their surrounding cloud or located near massive stars.\\
In this model, we do not account for the chemical effects induced by UV radiation originating from the protostar and its accretion region. The innermost particles in our simulation reach a minimum distance of 10 au from the protostar. We therefore assume that the contribution of UV radiation from the protostar can be neglected. To verify this assumption, we computed the ratio between the chemical rates for photo-reactions induced by the internal UV radiations and the external one.\\
The chemical rate for a photo-reaction is given by
\begin{equation}
k \propto F_\mathrm{UV} e^{-\gamma\ A_V}.
\end{equation}
Thus, we have
\begin{equation}
\frac{k_\mathrm{int}}{k_\mathrm{ext}} = \frac{F_{\mathrm{UV\ int}}}{F_{\mathrm{UV\ ext}}}\ 
e^{\gamma (A_V^{\mathrm{ext}} - A_V^{\mathrm{int}})}
\label{eq:k_int_k_ext}
\end{equation}
and
\begin{equation}
\label{eq:FUV}
F_{\mathrm{UV\ int}}(r) = \frac{2\pi}{c^2} \left( \frac{R_\mathrm{acc}}{r} \right)^2
\int_{\nu_1}^{\nu_2} \frac{\nu^2}{e^{(h\nu / kT)} - 1} \mathrm{d}\nu
\end{equation}
where $k_\mathrm{int}$ and $k_\mathrm{ext}$ denote the chemical rates driven by the internal UV field (originating from the protostar and its accretion region) and by the external interstellar radiation field, respectively. The interstellar UV flux is taken as $F_{\mathrm{UV\ ext}} = 10^8~\mathrm{photons~cm^{-2}~s^{-1}}$. $A_V^{\mathrm{ext}}$ corresponds to the visual extinction between the particle and the interstellar medium, while $A_V^{\mathrm{int}}$ is the visual extinction between the particle and the protostar. $F_{\mathrm{UV\ int}}$ is the UV flux from the protostar and its accretion region where $R_\mathrm{acc}$ denotes the radius of the accretion region at the protostellar surface. $\gamma$ is the UV attenuation parameter, it converts the visual extinction $A_V$ into UV extinction\\
We assume an inner gap of 1 au between the envelope and the protostar. The calculation was performed for the closest particle at the latest stage of our model, corresponding to the faintest envelope and the highest $F_{\mathrm{UV\ int}}$ value. Using equations 60 to 63 of \cite{Marchand2025a} (and adapting $r$ and $r_{\max}$), we obtain $A_V^{\mathrm{ext}} = 60~\mathrm{mag}$ and $A_V^{\mathrm{int}} = 157~\mathrm{mag}$. Equation \ref{eq:FUV} gives
\[
\frac{F_{\mathrm{UV\ int}}}{F_{\mathrm{UV\ ext}}} = 8.97 \times 10^6.
\]
Assuming $\gamma = 2$, Eq. \ref{eq:k_int_k_ext} gives: 
\[
\frac{k_\mathrm{int}}{k_\mathrm{ext}} = 5 \times 10^{-78}.
\]
Even under these conditions, we find that $k_\mathrm{int} \ll k_\mathrm{ext}$. This conclusion holds for $\gamma > 0.2$, which applies to all photo-reactions included in our network ($\gamma$ values range from 0.5 to 4). We therefore conclude that UV radiation from the protostar can safely be neglected in our model. However, it may become important at later evolutionary stages, when the envelope is significantly thinner, or in regions located closer to the protostar.

\section{Calculation of the temperature for the quiescent and outburst models}
\label{app:eq1&2}
The calculation of the temperature in the quiescent case is explained in \citealt{Marchand2025a} (see their Sect. 2.2.4 and their equation 24). As this demonstration is also important to derive the temperature in the case of outbursts, we provide it here as well.\\
In APE, the protostar is considered to behave like a blackbody with a temperature of $T_*$. Thus, the Stefan-Boltzmann law expresses the radiative flux emitted from the protostar’s surface per unit area, $\Phi_e (R_*)$:
\begin{equation}
    \label{eq:f1}
    \Phi_e (R_*) = \sigma T_*^4 ,
\end{equation}
where $R_*$ is the radius of the protostar and $\sigma$ is the Stefan–Boltzmann constant. Assuming that the envelope is optically thin to the infrared emission of the protostar, which is considered as a black body, the total energy flux is conserved when integrated over a sphere. Therefore, the link between the flux emitted by the protostar and the flux received in the envelope at a distance $r$ from the protostar, $\Phi_r (r)$, is:
\begin{equation}
    \label{eq:f2}
    4 \pi r^2 \Phi_r (r) = 4 \pi R_*^2 \Phi_e (R_*) .
\end{equation}
Furthermore, the cloud in which our system is located has a temperature of $T_{mc} = 10$ K. Consequently, \citealt{Marchand2025a} added a term related to the blackbody radiation associated with a medium at temperature $T_{mc}$:
\begin{equation}
    \label{eq:f3}
    \Phi_r (r) = \left(\frac{R_*}{r}\right)^2 \Phi_e (R_*) + \sigma T_{mc}^4 .
\end{equation}
Then, they assumed that the dust grains making up the protostellar envelope also behave as black bodies, meaning that the emitted and received fluxes are equal:
\begin{equation}
    \label{eq:f4}
    \Phi_{grain} (r) = \Phi_r (r) = \sigma T_{grain} (r)^4 .
\end{equation}
They also assume that thermal equilibrium between the grains and the surrounding gas is reached quickly. They therefore considered that
\[
    T (r) = T_{grain} (r) .
\]
We thus retrieve equation \ref{eq:quiescent_temperature} (equation 24 in \citealt{Marchand2025a}), where $T_\mathrm{quiescent} (r)$ is equal to $T (r)$:
\begin{equation}
    \label{eq:f5}
    T_\mathrm{quiescent} (r) = T (r) = \left[\left(\frac{R_*}{r}\right)^2 T_*^4 + T_{mc}^4\right]^\frac{1}{4} .
\end{equation}
\\
In the outburst case, we multiplied the radiative flux emitted by the protostar by a factor $f$. Therefore, \ref{eq:f3} becomes:
\begin{equation}
    \label{eq:f6}
    \Phi_r (r) = \left(\frac{R_*}{r}\right)^2 f \Phi_e (R_*) + \sigma T_{mc}^4 .
\end{equation}
and we retrieved equation \ref{eq:outburst_temperature}:
\begin{equation}
    \label{eq:f7}
    T_\mathrm{outburst} (r) = \left[f \left(\frac{R_*}{r}\right)^2 T_*^4 + T_{mc}^4\right]^\frac{1}{4} .
\end{equation}

\clearpage
\section{Model-observation comparison for COMs}
\label{app:model_observation_comparison}
\begin{figure}[h]
    \begin{minipage}[c]{\textwidth}
    \centering
    \includegraphics[width=1\textwidth]{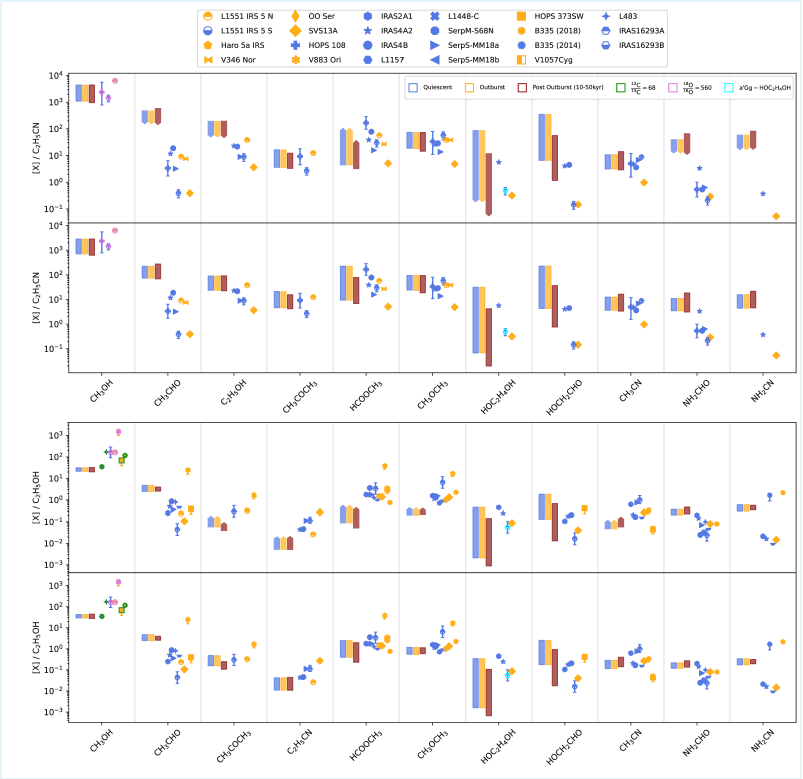}
    \caption{Same as Fig. \ref{fig:comparison_obs} but with different species of reference.}
    \label{fig:comp_obs_first}
    \end{minipage}
\end{figure}

\begin{figure*}[h]
    \centering
    \includegraphics[width=1\textwidth]{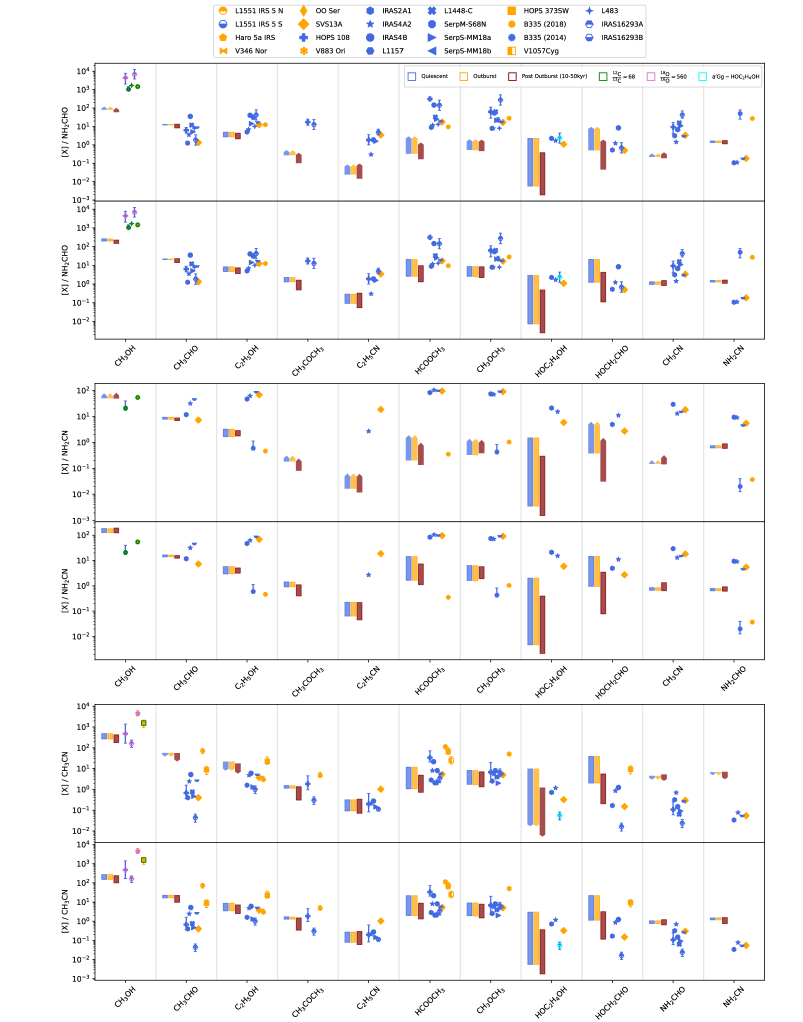}
    \addtocounter{figure}{-1}
    \caption{Continuation of Fig. \ref{fig:comp_obs_first}}
\end{figure*}
\begin{figure*}[h]
    \centering
    \includegraphics[width=1\textwidth]{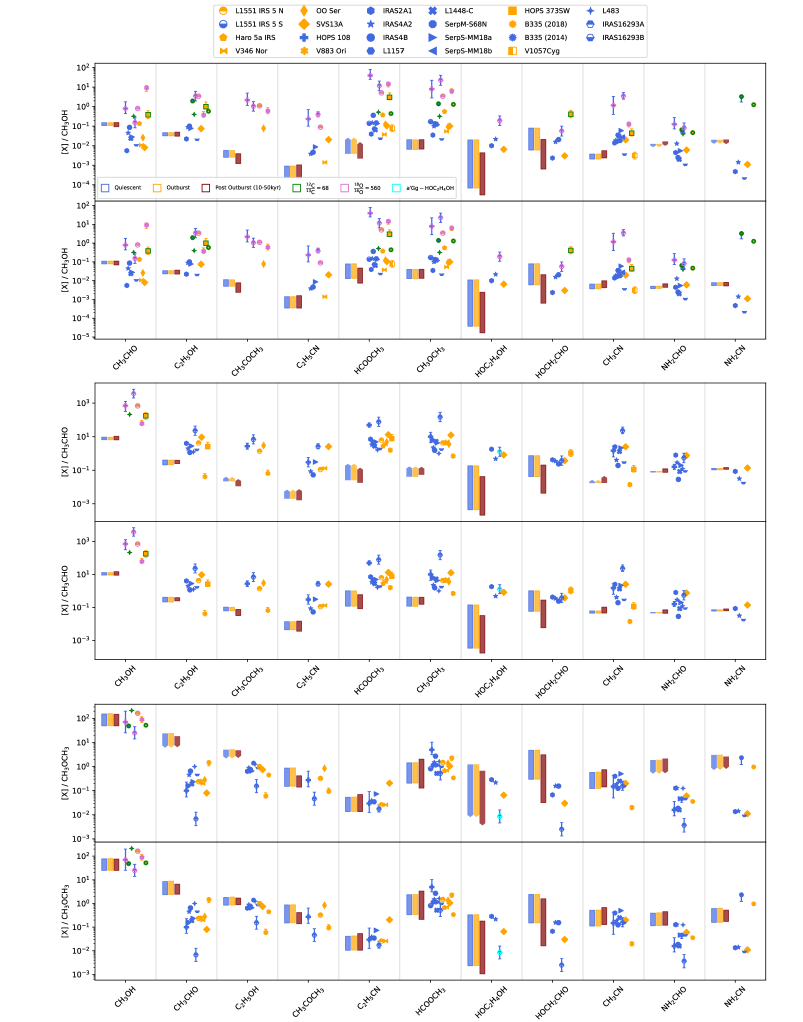}
    \addtocounter{figure}{-1}
    \caption{Continuation of Fig. \ref{fig:comp_obs_first}}
\end{figure*}
\begin{figure*}[h]
    \centering
    \includegraphics[width=1\textwidth]{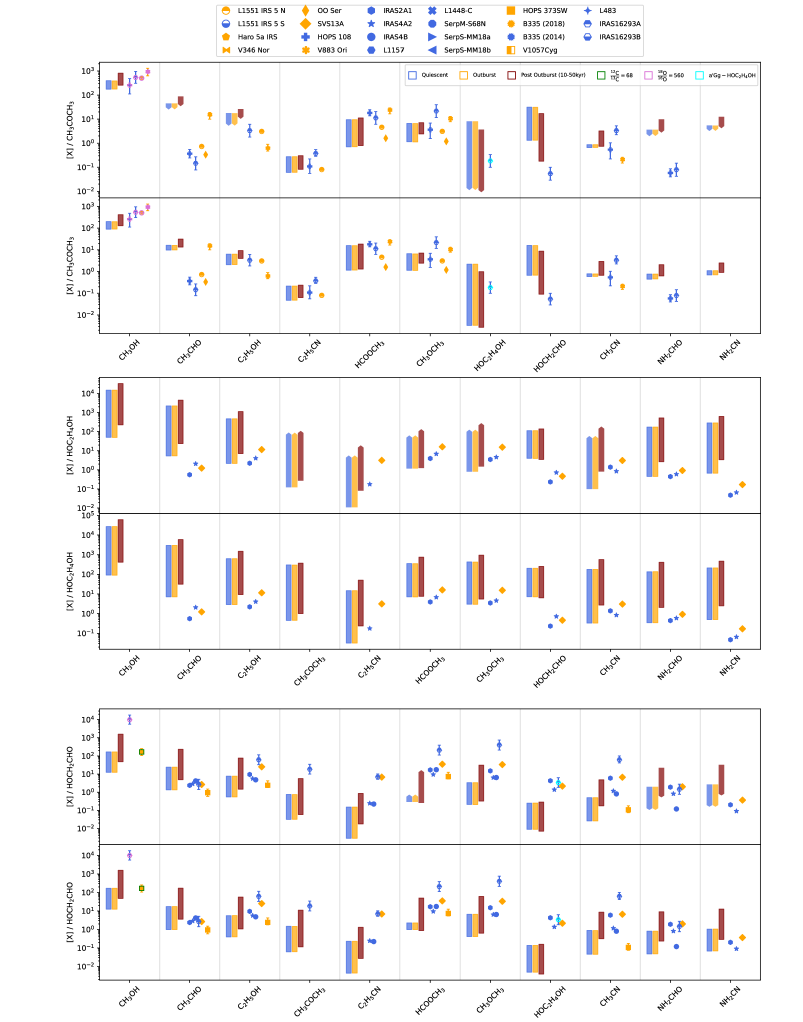}
    \addtocounter{figure}{-1}
    \caption{Continuation of Fig. \ref{fig:comp_obs_first}}
    \label{fig:comp_obs_last}
\end{figure*}

\end{document}